\documentclass[11pt,a4paper]{article}
\pdfoutput=1
\usepackage{jheppub}
\usepackage[T1]{fontenc}
\usepackage{lmodern,slashed,booktabs,longtable,array,xcolor,mathtools}
\usepackage{needspace,placeins}
\allowdisplaybreaks[2]
\newcommand{\MS}{\overline{\mathrm{MS}}}
\newcommand{\ep}{\epsilon}
\newcommand{\dd}{\mathrm{d}}
\newcommand{\ii}{\mathrm{i}}
\newcommand{\Li}{\operatorname{Li}}
\newcommand{\Cl}{\operatorname{Cl}_2}

\newcommand{\ringzeta}{\mathring\zeta_\psi}
\title{Off-lightcone Wilson-line operators in gradient flow at next-to-next-to-leading order}
\author[a]{Xiang-Peng~Wang}
\affiliation[a]{Institute of Particle Physics and Key Laboratory of Quark and Lepton Physics (MOE),\\
Central China Normal University, Wuhan, Hubei 430079, China}
\emailAdd{xpwang@ccnu.edu.cn}
\abstract{We consider gauge-invariant off-lightcone Wilson-line operators
  constructed from local fields connected by spacelike or timelike Wilson lines.
  The lattice evaluation of their matrix elements requires
  a suitable renormalization prescription. Gradient flow offers such
  a prescription and, by smoothing
  the fields, suppresses ultraviolet fluctuations and can reduce statistical
  noise. We use the previously established matching framework based on
  the one-dimensional auxiliary-field formalism. In the small-flow-time
  limit, this framework reduces the matching of nonlocal off-lightcone
  Wilson-line operators to that of local current operators.
  We calculate the matching coefficients for these local current operators
  from the gradient-flow scheme to the modified minimal subtraction
  ($\MS$) scheme through next-to-next-to-leading order (NNLO) in QCD.
  Our results can be applied to lattice calculations of unpolarized and
  polarized quark and gluon quasi-parton distributions, as well as meson
  and baryon quasi-distribution amplitudes.
  They also apply to gluonic correlators entering quarkonium transport
  and heavy-quark momentum diffusion in the quark-gluon plasma (QGP).
  Further applications include quarkonium decay and production in
  potential nonrelativistic QCD (pNRQCD) and
  spin-dependent relativistic corrections to the static potential.}
\begin{document}
\maketitle

\section{Introduction}\label{sec:intro}
The off-lightcone Wilson-line operators considered here are
gauge-invariant nonlocal operators constructed from local fields
connected by spacelike or timelike Wilson lines. These operators enter
a broad range of QCD observables accessible to lattice calculations.
For spacelike cases, their hadronic matrix elements provide access to lightcone
parton distribution functions (PDFs) through quasi-PDFs in large-momentum effective theory (LaMET),
pseudo-PDFs, and related lattice-cross-section formulations
\cite{Ji:2013dva,Ji:2014gla,Ma:2014lattice,Radyushkin:2017cyf,Ma:2017crosssections,Izubuchi:2018srq}.
For a review, see ref.~\cite{Ji:2020ect} and references therein.
For timelike cases, the relevant gluonic correlators can be related
to the long-distance matrix elements (LDMEs) entering nonrelativistic QCD
(NRQCD) factorization~\cite{Bodwin:1994jh} for quarkonium decay and
production. This connection has been established within potential NRQCD
(pNRQCD)~\cite{Pineda:1997bj,Brambilla:1999xf}; see
refs.~\cite{Brambilla:2001xy,Brambilla:2002nu,Brambilla:2020decay,Brambilla:2020ojz,Brambilla:2021abf,Brambilla:2022rjd,Brambilla:2022ayc}.
Field-strength insertions in Wilson loops
also determine the static force and relativistic corrections to the static
potential~\cite{Brambilla:2000gk,Pineda:2000sz};
see ref.~\cite{Brambilla:2004jw} for a review.
Thermal gluonic correlators also enter
the description of heavy-quark momentum diffusion
\cite{CasalderreySolana:2006diffusion,CaronHuot:2009thermalization}
and quarkonium transport
\cite{Brambilla:2017oqs,Brambilla:2018fireball,Yao:2021transport}
in the quark-gluon plasma (QGP); see ref.~\cite{Andronic:2024transportreview}
for a comparative review.

Evaluating these matrix elements on the lattice requires a suitable
renormalization prescription for the composite operators.
Gradient flow~\cite{Narayanan:2006rf,Luscher:2010iy,Luscher:2011bx,Luscher:2013cpa}
evolves fields to positive flow time, suppressing ultraviolet fluctuations.
It can reduce statistical noise in lattice observables
\cite{Asakawa:2013thermodynamics,Altenkort:2021hqd} and provides a
renormalization scheme whose conversion to the modified minimal subtraction
($\MS$) scheme can be computed perturbatively through the small-flow-time
expansion~\cite{Luscher:2011bx,Suzuki:2013gza}.
Matching coefficients and matrix elements of local flowed operators have
been studied perturbatively, with results available through
next-to-next-to-leading order (NNLO) for several classes of operators
\cite{Makino:2014taa,Harlander:2018zpi,Endo:2015iea,Hieda:2016lly,Borgulat:2023bca,Shindler:2023xpd,Harlander:2025twisttwo,Suzuki:2020zue,Harlander:2022tgk,Buhler:2023gsg,Rizik:2020naq,Mereghetti:2021nkt,Harlander:2022vgf,Crosas:2023anw,Crosas:2026ofx,Gui:2026flow,Boers:2020uqr,Harlander:2020duo,Harlander:2016vzb,Artz:2019bpr,Takaura:2025pao,Harlander:2025mass}.
Recent applications include determinations of parton-distribution
moments~\cite{Francis:2025pdf} and renormalized quark masses
\cite{Black:2025masses}, studies of charmonium mass decompositions and the trace anomaly
\cite{Bollweg:2026mass}, and the formulation of an operator product
expansion free of infrared renormalon ambiguities~\cite{Beneke:2025hlg}.

The use of gradient flow for quasi-parton distributions was proposed in
ref.~\cite{Monahan:2016bvm}, followed by one-loop matching at the
matrix-element level~\cite{Monahan:2017hpu}.
Further developments include baryon quasi-distribution amplitudes
\cite{Zhang:2026baryonflow} and renormalons in gradient-flow quasi-PDFs
\cite{Zhang:2025flowrenormalon}.
Related matching calculations have been performed for flowed chromoelectric and chromomagnetic
correlators~\cite{deLaCruz:2024nlo}.
The operator-level framework of ref.~\cite{Brambilla:2023vwm} uses the
one-dimensional auxiliary-field formalism
\cite{Samuel:1978iy,Gervais:1979fv}, reviewed in ref.~\cite{Dorn:1986dt},
to reduce the matching of off-lightcone Wilson-line operators to that of
local currents in the small-flow-time limit. This reduces the two-scale
matching problem to a single-scale calculation.

The one-loop matching coefficients~\cite{deLaCruz:2024nlo,Brambilla:2023vwm}
have been used in gradient-flow lattice calculations of heavy-quark
diffusion~\cite{Altenkort:2024bfield,Bollweg:2025hqdiff}.
The framework of ref.~\cite{Brambilla:2023vwm} has also been applied to the chromoelectric
correlator entering inclusive $P$-wave quarkonium decays
\cite{Brambilla:2026pwave}.
The same matching is relevant to spin-dependent potentials and
hybrid--quarkonium mixing studied with gradient flow
\cite{Eichberg:2024svw,Schlosser:2025tca}.
The static force has also been studied in gradient flow, both
perturbatively~\cite{Brambilla:2021egm} and on the
lattice~\cite{Brambilla:2024static}.
Lattice studies also address adjoint chromoelectric correlators
relevant to quarkonium dynamics in the QGP
\cite{Brambilla:2025adjoint}.

Using this established framework, we calculate the matching coefficients
for the local current operators from the gradient-flow scheme to the
$\MS$ scheme at NNLO in QCD. These coefficients provide the local-current
conversion factors at NNLO for the
corresponding off-lightcone Wilson-line operators.

The rest of the paper is organized as follows.
Section~\ref{sec:framework} introduces the gradient-flow formalism,
the local currents and off-lightcone Wilson-line operators, and their
renormalization and matching. Section~\ref{sec:calculation} describes the
NNLO calculation and presents the matching coefficients and their
numerical analysis. Section~\ref{sec:discussion} summarizes the results
and applications. Appendices~\ref{app:renormalization},
\ref{app:matching} and \ref{app:masters} collect the renormalization
constants and anomalous dimensions, field and vertex matching coefficients, and master integrals,
respectively.

\section{Theoretical framework}\label{sec:framework}
\subsection{QCD gradient flow}
We work throughout in Euclidean spacetime of dimension $D=4-2\ep$
and metric $\delta_{\mu\nu}$.
The terms spacelike and timelike refer to the corresponding
Minkowski-space Wilson lines.
Gradient flow evolves the QCD gauge field $A_\mu^a(x)$ and quark fields
$\psi_f^i(x)$ and $\bar\psi_f^i(x)$ into the flowed fields
$B_\mu^a(t,x)$, $\chi_f^i(t,x)$ and $\bar\chi_f^i(t,x)$ at flow time
$t\geq0$. Here $a$ and $i$ are adjoint and fundamental colour indices,
and $f$ labels the quark flavour. The fields satisfy the boundary conditions
\begin{equation}
 B_\mu^a(0,x)=A_\mu^a(x),\qquad
 \chi_f^i(0,x)=\psi_f^i(x),\qquad
 \bar\chi_f^i(0,x)=\bar\psi_f^i(x),
 \label{eq:flow-boundary}
\end{equation}
and the flow equations~\cite{Luscher:2010iy,Luscher:2011bx,Luscher:2013cpa}
\begin{align}
 \partial_t B_\mu^a
 &=\mathcal D_\nu^{ab}G_{\nu\mu}^b+\kappa \mathcal D_\mu^{ab}\partial_\nu B_\nu^b,
 \label{eq:flow}\\*
 \partial_t\chi_f
 &=\bigl[\mathcal D_\mu^{\rm F}\mathcal D_\mu^{\rm F}
          -\kappa(\partial_\mu B_\mu^a)T^a\bigr]\chi_f,
 \label{eq:flow-quark}\\*
 \partial_t\bar\chi_f
 &=\bar\chi_f\bigl[\overleftarrow{\mathcal D}_\mu^{\rm F}
       \overleftarrow{\mathcal D}_\mu^{\rm F}
          +\kappa(\partial_\mu B_\mu^a)T^a\bigr].
 \label{eq:flow-antiquark}
\end{align}
Here $T^a$ are anti-Hermitian colour generators in the fundamental
representation, with $[T^a,T^b]=f^{abc}T^c$ and
$\operatorname{tr}(T^aT^b)=-T_R\delta^{ab}$, where
$T_R=1/2$~\cite{Luscher:2010iy,Luscher:2011bx}.
The flowed field strength and covariant derivatives are
\begin{align}
 G_{\mu\nu}^a
 &=\partial_\mu B_\nu^a-\partial_\nu B_\mu^a
       +f^{abc}B_\mu^bB_\nu^c,
 \label{eq:flow-conventions}\\*
 \mathcal D_\mu^{ab}&=\delta^{ab}\partial_\mu-f^{abc}B_\mu^c,
 \label{eq:flow-adjoint-derivative}\\*
 \mathcal D_\mu^{\rm F}
 &=\partial_\mu+B_\mu^aT^a,
 \label{eq:flow-fundamental-derivative}\\*
 \overleftarrow{\mathcal D}_\mu^{\rm F}
 &=\overleftarrow\partial_\mu-B_\mu^aT^a.
 \label{eq:flow-left-derivative}
\end{align}
The flow-gauge parameter $\kappa$ drops out of gauge-invariant
observables. We set $\kappa=1$ and work in a general covariant $R_\xi$
gauge for QCD, keeping $\xi$ arbitrary in our perturbative calculations;
$\xi=1$ corresponds to Feynman gauge.

At positive flow time, the flowed gauge field requires no additional
field renormalization after the usual QCD renormalization, whereas flowed
quark fields require a multiplicative renormalization
\cite{Luscher:2011bx,Luscher:2013cpa}.
We adopt the ringed-fermion scheme~\cite{Makino:2014taa} and define
\begin{equation}
 \mathring\chi_f(t,x)=\mathring Z_\chi^{1/2}(t)\chi_{f,0}(t,x),\qquad
 \mathring{\bar\chi}_f(t,x)=\mathring Z_\chi^{1/2}(t)\bar\chi_{f,0}(t,x).
 \label{eq:ringed-fields}
\end{equation}
The subscript $0$ denotes bare quantities. The normalization is fixed by
\begin{equation}
 \left.\mathring Z_\chi(t)
 \left\langle\sum_{f=1}^{n_f}\bar\chi_{f,0}(t,x)
 \overleftrightarrow{\slashed{\mathcal D}}^{\rm F}\chi_{f,0}(t,x)\right\rangle
 \right|_{m_f=0}
 =-\frac{2N_c n_f}{(4\pi)^2t^2}.
 \label{eq:ringed}
\end{equation}
Here $\overleftrightarrow{\mathcal D}_\mu^{\rm F}
 =\mathcal D_\mu^{\rm F}-\overleftarrow{\mathcal D}_\mu^{\rm F}$;
$N_c$ is the number of colours and $n_f$ is the number of light quark
flavours. The quark mass is denoted by $m_f$; we neglect quark masses
throughout this study.
Gauge-invariant local composite operators constructed from flowed gauge
fields and ringed quark fields at positive flow time are ultraviolet
finite after the usual QCD renormalization~\cite{Luscher:2011bx,Luscher:2013cpa,Makino:2014taa}.
The conversion from $\MS$-renormalized flowed quark fields to ringed quark fields
has been computed at one loop~\cite{Makino:2014taa} and two
loops~\cite{Artz:2019bpr}. We collect these results in
appendix~\ref{app:renormalization}.

For perturbative calculations, the flow equations can be implemented in
a local \mbox{$(D+1)$-dimensional} formulation, with Lagrange-multiplier fields
that generate flow lines and flow vertices. We use this formulation;
its Lagrangian and Feynman rules can be found in
refs.~\cite{Luscher:2011bx,Luscher:2013cpa,Artz:2019bpr}.

\subsection{Renormalization and matching}
The off-lightcone Wilson-line operators considered here are
\begin{align}
 \mathcal O_\Gamma(zv)&=\bar\psi(zv)\Gamma W(zv,0)\psi(0),
 \label{eq:operator}
 \\
 \mathcal O^{\mu\nu\alpha\beta}(zv)
 &=F^{\mu\nu}(zv)W(zv,0)F^{\alpha\beta}(0),
 \label{eq:gluon-operator}
\end{align}
and operators with chromomagnetic field insertions into Wilson lines or
loops. The straight path-ordered Wilson line is defined by
\begin{equation}
 W(zv,0)=\mathcal P\exp\left[-\int_0^z\dd s\,
 v\cdot A(sv)\right].
 \label{eq:wilson}
\end{equation}
Here $v$ is a Euclidean unit vector, $v^2=1$, and $z$ is real.
For convenience, we choose $z>0$ in the following discussion.
The Wilson line is in the fundamental representation for the quark
operator; in the gluon operator it contracts the adjoint colour indices
of the two field strengths.
For simplicity, colour indices are suppressed where no confusion arises.
Here $F^{\mu\nu}$ denotes the QCD
field strength and $G^{\mu\nu}$ its flowed counterpart.\footnote{Following the gauge-field normalization of refs.~\cite{Luscher:2010iy,Luscher:2011bx},
we absorb the gauge coupling into the gauge field. Thus
$F^{\mu\nu}$ here corresponds to $gF^{\mu\nu}$ in
refs.~\cite{Brambilla:2023vwm,Ji:2020ect}, and eq.~\eqref{eq:gluon-operator}
contains no additional factor of $g^2$.}
$\Gamma$ specifies the Dirac structure of the quark operator.
Flavour labels are suppressed.

The gluon components parallel and transverse to the Wilson-line
direction renormalize differently~\cite{Braun:2020ymy}. We introduce the
projectors
\begin{equation}
 \delta_\parallel^{\mu\nu}=v^\mu v^\nu,\qquad
 \delta_\perp^{\mu\nu}=\delta^{\mu\nu}-v^\mu v^\nu,
 \label{eq:tensor-components}
\end{equation}
and the projected field strengths
\begin{equation}
 F_{\parallel\perp}^{\mu\nu}
   =\delta_\parallel^{\mu\alpha}\delta_\perp^{\nu\beta}F^{\alpha\beta},
 \qquad
 F_{\perp\perp}^{\mu\nu}
   =\delta_\perp^{\mu\alpha}\delta_\perp^{\nu\beta}F^{\alpha\beta}.
 \label{eq:projected-field-strengths}
\end{equation}
For a timelike Wilson line, these correspond to chromoelectric and
chromomagnetic components. The projected gluon bilocal operators are
\begin{equation}
 \mathcal O_{rs}^{\mu\nu\alpha\beta}(zv)
 =F_r^{\mu\nu}(zv)W(zv,0)F_s^{\alpha\beta}(0),
 \qquad r,s\in\{\parallel\perp,\perp\perp\}.
 \label{eq:projected-gluon-operators}
\end{equation}
For $r=s$, we write $\mathcal O_r\equiv\mathcal O_{rr}$.

In the auxiliary-field formalism, the Lagrangian in Euclidean spacetime is
\begin{equation}
 \mathcal L_{h_v}=\bar h_v(v\cdot D)h_v,
 \label{eq:auxiliary-lagrangian}
\end{equation}
where $h_v$ is an auxiliary Grassmann scalar field in the fundamental
($\mathrm F$) or adjoint ($\mathrm A$) representation, and
$D^\mu=\partial^\mu+A^\mu$ is the QCD covariant derivative in that
representation. A mass counterterm $\delta m\,\bar h_vh_v$
subtracts the linearly divergent part of the Wilson-line self-energy.
In the unflowed theory, the perturbative mass correction vanishes in
dimensional regularization. It is nonzero with a lattice cutoff or in the
gradient-flow scheme. The one-loop result for
$\delta m(t)$ in the gradient-flow scheme was obtained in
ref.~\cite{Brambilla:2023vwm}. For the adjoint representation, its linearly
divergent part was extracted from flowed Polyakov loops in a quenched lattice
calculation~\cite{Brambilla:2026pwave}.
We treat $\delta m$ as an independent input and do not calculate it here.

At nonzero separation, the quark and projected gluon bilocal operators renormalize
multiplicatively~\cite{Ishikawa:2017faj,Ji:2017oey,Zhang:2018diq},
assuming a regularization that preserves chiral symmetry in the quark case:
\begin{align}
 \mathcal O_\Gamma^{\rm B}(zv,\Lambda)
 &=e^{\delta m^{\mathrm F}(\Lambda)z}Z_{J_q}^{\,2}(\Lambda,\mu)
 \mathcal O_\Gamma^{\MS}(zv,\mu),
 \label{eq:quark-operator-renormalization}\\
 \mathcal O_{rs}^{\rm B}(zv,\Lambda)
 &=e^{\delta m^{\mathrm A}(\Lambda)z}Z_{J_r}(\Lambda,\mu)
 Z_{J_s}(\Lambda,\mu)\mathcal O_{rs}^{\MS}(zv,\mu).
 \label{eq:operator-renormalization}
\end{align}
Lorentz indices are suppressed where no confusion can arise. The superscript B denotes bare operators,
$\Lambda$ is the ultraviolet cutoff, and $\mu$ is the renormalization scale.
The factors $Z_{J_q}$ and $Z_{J_r}$ account for logarithmic
divergences and are the $\MS$ renormalization constants
for the local currents
\begin{equation}
 J_q=\bar\psi h_v^{\mathrm F},\qquad
 J_{\parallel\perp}^{\mu\nu}
   =F_{\parallel\perp}^{\mu\nu}h_v^{\mathrm A},\qquad
 J_{\perp\perp}^{\mu\nu}
   =F_{\perp\perp}^{\mu\nu}h_v^{\mathrm A}.
 \label{eq:gluon-currents}
\end{equation}
In physical matrix elements, these currents renormalize as
\begin{equation}
 J_q^{\rm B}=Z_{J_q}J_q^{\MS},\qquad
 J_r^{\rm B}=Z_{J_r}J_r^{\MS},
 \qquad r\in\{\parallel\perp,\perp\perp\}.
 \label{eq:local-current-renormalization}
\end{equation}

For each current, our anomalous-dimension
convention is
\begin{equation}
 \gamma_J=\mu\frac{\dd\ln(Z_J)}{\dd\mu},\qquad
 \mu\frac{\dd J^{\MS}}{\dd\mu}=-\gamma_JJ^{\MS}.
 \label{eq:current-anomalous-dimension-convention}
\end{equation}
For the gluon currents, we also write $Z_r\equiv Z_{J_r}$ and
$\gamma_r\equiv\gamma_{J_r}$, as in appendix~\ref{app:renormalization}.

Chromomagnetic insertions into fundamental Wilson lines or loops are
described by the additional current
\begin{equation}
 J_F
   =\bar h_v^{\mathrm F}F_{\perp\perp}h_v^{\mathrm F},
 \qquad J_F^{\rm B}=Z_FJ_F^{\MS}.
 \label{eq:fundamental-magnetic-current}
\end{equation}
Its matching coefficient below is defined at zero momentum transfer
through the insertion.

To determine the finite matching coefficients, we replace the quark fields
by ringed fields and $F^{\mu\nu}$ by $G^{\mu\nu}$.
All gauge fields directly coupled to
$h_v$ are evaluated at the same flow
time $t$, while $h_v$ itself is not flowed. 
The small-flow-time expansions of these currents in physical matrix elements are
\begin{align}
 J_q(t)&=c_{\psi h_v}(t,\mu)J_q^{\MS}(\mu)+O(t),
 \label{eq:local-matching}\\
 J_{\parallel\perp}(t)&=c_{\parallel\perp}(t,\mu)J_{\parallel\perp}^{\MS}(\mu)+O(t),
 \label{eq:local-matching-parallel}\\
 J_{\perp\perp}(t)&=c_{\perp\perp}(t,\mu)J_{\perp\perp}^{\MS}(\mu)+O(t),
 \label{eq:local-matching-perp}\\
 J_F(t)&=c_F(t,\mu)J_F^{\MS}(\mu)+O(t).
 \label{eq:local-matching-F}
\end{align}
Here the currents on the left-hand side are the flowed currents, and $O(t)$
denotes higher-dimension operators. The field and vertex matching factors
combine as
\begin{align}
 c_{\psi h_v}&=\ringzeta\,\zeta_{h_v}^{\mathrm F}\,\zeta_{\psi h_v},
 \label{eq:three-factor}\\
 c_{\parallel\perp}
   &=\zeta_A\zeta_{h_v}^{\mathrm A}\zeta_{\parallel\perp},
 \label{eq:gluon-factorization-parallel}\\
 c_{\perp\perp}
   &=\zeta_A\zeta_{h_v}^{\mathrm A}\zeta_{\perp\perp},
 \label{eq:gluon-factorization-perp}\\
 c_F&=\zeta_A(\zeta_{h_v}^{\mathrm F})^2\zeta_F.
 \label{eq:gluon-factorization}
\end{align}
Here $\ringzeta$, $\zeta_A$ and $\zeta_{h_v}$ match the ringed quark,
gauge and mass-subtracted auxiliary fields, respectively; the remaining
factors match the amputated current vertices after external-field matching.

At fixed nonzero $z$, with $\sqrt{8t}\ll z$ and fixed external scales,
the local expansion applies to each current separately. Products of
renormalized currents at distinct points require no additional
renormalization. After mass subtraction, the bilocal matching
coefficient is therefore the product of the local-current matching coefficients
at leading order in the small-flow-time expansion. Thus
\begin{align}
 \mathcal O_\Gamma(t,zv)
 &=e^{\delta m^{\mathrm F}(t)z}
 \left[\mathcal C_\psi(t,\mu)\mathcal O_\Gamma^{\MS}(zv,\mu)+O(t)\right],
 \label{eq:bilocalmatching}\\
 \mathcal O_{rs}(t,zv)
 &=e^{\delta m^{\mathrm A}(t)z}
 \left[\mathcal C_{rs}(t,\mu)\mathcal O_{rs}^{\MS}(zv,\mu)+O(t)\right],
 \label{eq:gluon-bilocal-matching}
\end{align}
with
\begin{align}
 \mathcal C_\psi&=c_{\psi h_v}^{\,2},
 \label{eq:bilocal-coefficient}\\
 \mathcal C_{rs}&=c_r c_s,\qquad r,s\in\{\parallel\perp,\perp\perp\}.
 \label{eq:gluon-bilocal-coefficients}
\end{align}
For equal projections, $\mathcal C_r\equiv\mathcal C_{rr}=c_r^2$;
each chromomagnetic insertion on a fundamental Wilson line instead
contributes one factor of $c_F$.

The operators used for unpolarized and helicity gluon quasi-PDFs can be
written as contractions or linear combinations of the projected bilocal operators
in eq.~\eqref{eq:projected-gluon-operators}~\cite{Ji:2020ect,Zhang:2018diq}.
Their matching follows from eq.~\eqref{eq:gluon-bilocal-coefficients},
with each projected operator matched separately.

\section{Computation and results}\label{sec:calculation}
We expand in $a_s=\alpha_s(\mu)/(4\pi)$, where $\alpha_s=g^2/(4\pi)$, and use
$L=\ln(2\mu^2t e^{\gamma_E})$ and
$\widetilde\mu^2=\mu^2 e^{\gamma_E}/(4\pi)$, where $\gamma_E$ is
the Euler--Mascheroni constant.
The colour factors are $C_F=(N_c^2-1)/(2N_c)$ and $C_A=N_c$.

\subsection{Projection method}\label{sec:projection}
We extract the matching coefficients from amputated Green functions using
the method of projectors~\cite{Gorishny:1983ope,Gorishny:1987ope}, as
applied to flowed operators in refs.~\cite{Harlander:2018zpi,Borgulat:2023bca}.
The Green functions are Taylor expanded in the external momenta before
loop integration, and the projectors are normalized to the corresponding
tree-level amplitudes. For the quark current we
project onto its tree-level Dirac structure. The field matching coefficients
are extracted from the corresponding two-point functions.

For the gluon currents, we differentiate the vertex with respect to the
external gluon momentum $q^\sigma$ and set $q=0$. After dividing out the
tree-level colour factor and a common factor of $\ii$, the tree-level
tensors following from eq.~\eqref{eq:projected-field-strengths} are
\begin{align}
 \mathcal T_{\parallel\perp}^{\mu\nu;\rho\sigma}
 &=v^\mu\bigl(v^\sigma\delta_\perp^{\nu\rho}
               -v^\rho\delta_\perp^{\nu\sigma}\bigr),
 \label{eq:gluon-tree-tensor-parallel}\\*
 \mathcal T_{\perp\perp}^{\mu\nu;\rho\sigma}
 &=\delta_\perp^{\mu\sigma}\delta_\perp^{\nu\rho}
   -\delta_\perp^{\nu\sigma}\delta_\perp^{\mu\rho}.
 \label{eq:gluon-tree-tensors}
\end{align}
We use the normalized Lorentz projectors
\begin{align}
 \mathcal P_{\perp\perp}^{\mu\nu;\rho\sigma}
 &=\frac{\delta_\perp^{\mu\sigma}\delta_\perp^{\nu\rho}
             -\delta_\perp^{\nu\sigma}\delta_\perp^{\mu\rho}}
          {2(D-1)(D-2)},
 \label{eq:gluon-projector-perp}\\*
 \mathcal P_{\parallel\perp}^{\mu\nu;\rho\sigma}
 &=-\frac{v^\mu\delta_\perp^{\nu\sigma}v^\rho}{D-1}.
 \label{eq:gluon-projectors}
\end{align}
Here $\mu$ and $\nu$ are the operator indices, $\rho$ is the Lorentz index
of the external gluon, and $\sigma$ labels the momentum derivative.
The projectors satisfy
$\mathcal P_r^{\mu\nu;\rho\sigma}\mathcal T_r^{\mu\nu;\rho\sigma}=1$.
The normalization factors follow from $\operatorname{tr}\delta_\perp=D-1$
and $\mathcal T_{\perp\perp}^{\mu\nu;\rho\sigma}
\mathcal T_{\perp\perp}^{\mu\nu;\rho\sigma}=2(D-1)(D-2)$.
The $\parallel\perp$ projector
projects onto $v^\mu v^\rho\delta_\perp^{\nu\sigma}$ and
annihilates $v^\mu\delta_\perp^{\nu\rho}v^\sigma$. For example, when
the momentum transfer through the current vanishes, the latter structure
also arises from the gauge-noninvariant operator
$v^\mu A_\perp^\nu(v\cdot D+\delta m^{\mathrm A})h_v^{\mathrm A}$,
with the mass counterterm as defined in
section~\ref{sec:framework}~\cite{Brambilla:2023vwm}.
Here $A_\perp^\nu=\delta_\perp^{\nu\rho}A^\rho$.
Momentum derivatives act on the complete integrand, including flow
kernels and propagators.

\subsection{Calculation and matching}\label{sec:procedure}
\begin{figure}[tbp]
 \centering
 \includegraphics{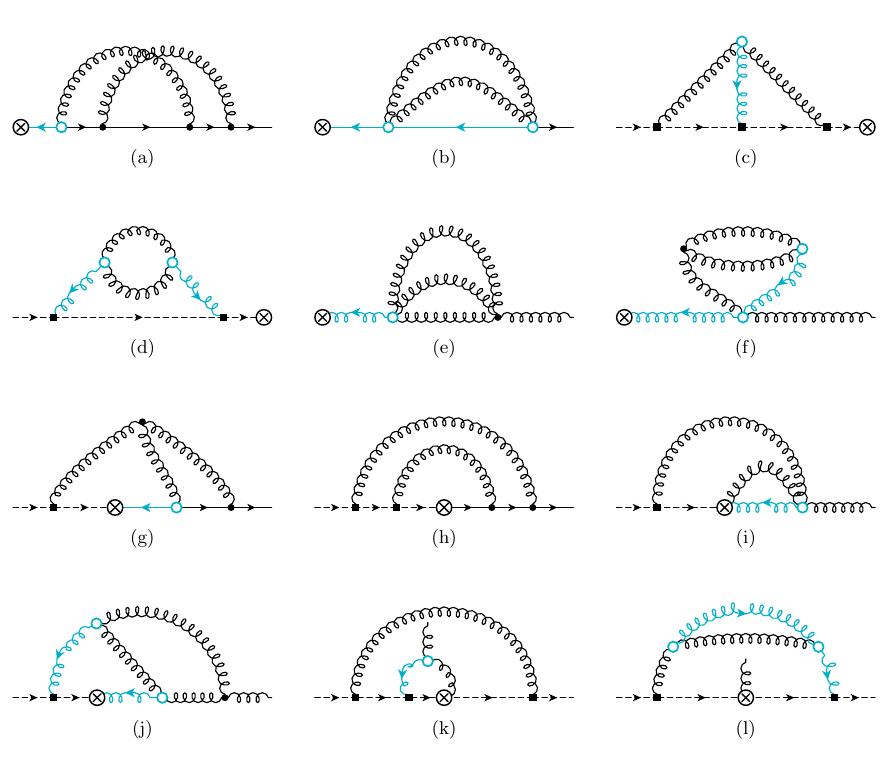}
 \caption{Representative two-loop diagrams entering the field matching
 for $\psi$ (a,b), $h_v$ (c,d) and $A^\mu$ (e,f), and the current
 matching for $J_q$ (g,h), $J_{\parallel\perp}$ and $J_{\perp\perp}$
 (i,j), and $J_F$ (k,l). Solid, dashed and curly lines denote light
 quarks, auxiliary fields and gluons, respectively. Propagators, including flowed propagators, are black, while retarded flow
 lines are cyan, with arrows towards increasing flow time. Open cyan circles mark flow vertices, filled
 dots ordinary QCD vertices, and filled squares auxiliary-field--gluon
 vertices, whose gluon fields are evaluated at flow time $t$. Crossed
 circles denote the insertions of $\chi(t)$, $h_v$ and $B^\mu(t)$
 in the field diagrams, and the flowed local currents in the vertex diagrams.}
 \label{fig:representative-diagrams}
\end{figure}

We generate diagrams with QGRAF~\cite{Nogueira:1991ex} and use
FORM~\cite{Ruijl:2017dtg} for the Dirac and colour algebra and the
projections described above. After extracting the required momentum
structures, all external momenta are set to zero. The resulting
integrals contain massless and static propagators, exponential factors
from the flow kernels, and integrations over internal flow times. Integration-by-parts
identities~\cite{Chetyrkin:1981ibp}, including derivatives of the flow factors, and flow-time
identities~\cite{Artz:2019bpr,Lange:2021thesis} reduce these integrals to the
master integrals listed in appendix~\ref{app:masters}. We retain the flow-time boundary
terms and solve the reduction systems
with Kira~\cite{Maierhoefer:2017kira,Klappert:2020kira}.
The master integrals are evaluated analytically.
Representative two-loop diagrams are shown in
figure~\ref{fig:representative-diagrams}.

Ultraviolet divergences are removed by the coupling, gauge-parameter,
field and current renormalization constants collected in
appendix~\ref{app:renormalization}, keeping $\xi$ arbitrary.
The ringed-field conversion supplies the finite normalization of the
flowed quark fields. The remaining infrared divergences cancel when matching
to the unflowed matrix elements, with the same static-propagator
prescription $-\ii/(v\cdot k-\ii0^+)$ on both sides, where $k$ is the
momentum carried by the auxiliary field.
In the massless, zero-external-momentum limit, the unflowed integrals
are scaleless and vanish in dimensional regularization: their ultraviolet
and infrared poles cancel. The infrared subtraction is therefore fixed
by the corresponding unflowed operator and external-field
renormalization constants~\cite{Brambilla:2023vwm,Borgulat:2023bca}.
After these subtractions, we obtain the finite field and vertex matching
coefficients and combine them according to
eqs.~\eqref{eq:three-factor}--\eqref{eq:gluon-factorization} to give the
matching coefficients of the local currents.

\subsection{Results}\label{sec:explicit-results}
We give the matching coefficients for the local currents in the conventions of
section~\ref{sec:framework}, at leading order in the small-flow-time
expansion. The separate field and vertex factors are collected in
appendix~\ref{app:matching}. The functions $\Li_n(x)$ are
polylogarithms, $\zeta(n)$ is the Riemann zeta function, and
$\Cl(\theta)=\operatorname{Im}\Li_2(e^{\ii\theta})$
is the Clausen function, with $\Cl(\pi/3)\simeq1.0149416064$.

\subsubsection{Quark current}
Combining the factors in eq.~\eqref{eq:three-factor} through $O(\alpha_s^2)$,
we obtain the matching coefficient $c_{\psi h_v}$, which is independent of $\xi$:
\begin{align}
 c_{\psi h_v}={}&1-a_sC_F\left[\frac32L+1+\frac12\ln(432)\right]\nonumber\\*
 &\hphantom{1}{}+a_s^2\Biggl\{L^2\left(\frac98C_F^2-\frac{11}{4}C_FC_A
 +C_FT_Rn_f\right)\nonumber\\*
 &\quad{}+L\biggl[C_F^2\left(\frac{11}{4}-\frac{4\pi^2}{3}+\frac34\ln(432)\right)
 +C_FC_A\left(\frac{\pi^2}{3}-\frac{31}{4}-\frac{11}{6}\ln(432)\right)
 \nonumber\\*
 &\qquad{}+C_FT_Rn_f\left(3+\frac23\ln(432)\right)\biggr]\nonumber\\*
 &\quad{}+C_F^2\biggl[2+\frac{4\pi^2}{3}\bigl[\ln(2)-1\bigr]
 -4\pi\operatorname{Cl}_2(\pi/3)+12\ln(2)\nonumber\\*
 &\qquad{}-\frac38\ln(3)\bigl[4+8\ln(2)+3\ln(3)\bigr]
 +\frac{11}{2}\Li_2(1/4)-\Li_3(1/4)-\frac43\zeta(3)\biggr]\nonumber\\*
 &\quad{}+C_FC_A\biggl[-\frac{239}{6}-\pi+\frac{\pi^2}{6}
 -\frac{\pi^2}{3}\ln(2)+\pi\operatorname{Cl}_2(\pi/3)
 -85\ln(2)+84\ln(3)\nonumber\\*
 &\qquad{}+2\ln^2(2)+17\Li_2(1/4)
 -\frac12\Li_3(1/4)-\frac23\zeta(3)\biggr]\nonumber\\*
 &\quad{}+C_FT_Rn_f\left[\frac{20}{3}+\frac{\pi^2}{3}\right]
 +\frac12C_{\chi,2}\Biggr\}+O(a_s^3).
 \label{eq:res-current-final}
\end{align}
Here $C_{\chi,2}$ is the two-loop constant at $L=0$ in the finite
conversion from $\MS$-renormalized flowed quark fields to ringed quark fields
in eq.~\eqref{eq:app-ringed}. The two-loop constant was computed in
ref.~\cite{Artz:2019bpr}; we use the numerical form quoted in
ref.~\cite{Borgulat:2023bca}:
\begin{equation}
 C_{\chi,2}\simeq-23.7947\,C_FC_A+30.3914\,C_F^2
                   -3.92255\,C_FT_Rn_f.
 \label{eq:cchi-numerical}
\end{equation}
For $N_c=3$, the current coefficient becomes
\begin{align}
 c_{\psi h_v}\simeq{}&1-a_s(2L+5.37895)\nonumber\\*
 &\hphantom{1}{}+a_s^2\bigl[\bigl(-9+\tfrac23n_f\bigr)L^2
       +(-72.7568+4.69708n_f)L\nonumber\\*
 &\hphantom{1+a_s^2\bigl[}{}-52.0835+5.33017n_f\bigr]+O(a_s^3).
 \label{eq:res-current-numerical}
\end{align}

\subsubsection{Gluon currents}\label{sec:gluon-results}
For the adjoint currents, we find
\begin{align}
 c_{\parallel\perp}={}&1+a_s^2 C_A^2
 \Biggl\{-\frac{2\pi^2}{3}L+\frac{2\pi^2}{3}\left(\ln(2)-\frac53\right)\nonumber\\*
 &\hphantom{1+a_s^2C_A^2\Biggl\{}{}-\frac{4\pi}{3}\left(-2+\sqrt3+\Cl(\pi/3)\right)\Biggr\}+O(a_s^3),
 \label{eq:gluon-adjoint-electric}\\
 c_{\perp\perp}={}&1+a_sC_A L\nonumber\\*
 &\hphantom{1}{}+a_s^2\Biggl\{L^2\left(\frac73C_A^2
       -\frac23C_AT_Rn_f\right)
       +L\left[C_A^2\left(\frac{34}{9}-\frac{2\pi^2}{3}\right)
       -\frac{26}{9}C_AT_Rn_f\right]\nonumber\\*
 &\quad{}+C_A^2\biggl[\frac{196}{27}+\pi-\frac{\pi^2}{6}
       -\frac{4\pi}{3}\Cl(\pi/3)
       \nonumber\\*
 &\qquad{}+\left(72+\frac{2\pi^2}{3}\right)\ln(2)
       -\frac{10}{3}\ln^2(2)-\frac{123}{2}\ln(3)
       +4\Li_2(1/4)\biggr]
       \nonumber\\*
 &\quad{}-\frac{C_AT_Rn_f}{27}\left(89+3\pi^2\right)\Biggr\}+O(a_s^3).
 \label{eq:gluon-adjoint-magnetic}
\end{align}
For the chromomagnetic insertion on a fundamental Wilson line, we obtain
\begin{align}
 c_F={}&1+a_sC_A L\nonumber\\*
 &\hphantom{1}{}+a_s^2\Biggl\{L^2\left(\frac73C_A^2
       -\frac23C_AT_Rn_f\right)
       +L\left(\frac{34}{9}C_A^2-\frac{26}{9}C_AT_Rn_f\right)
       \nonumber\\*
 &\quad{}+C_A^2\biggl[\frac{196}{27}+\pi+\frac{\pi^2}{2}
       +72\ln(2)-\frac{10}{3}\ln^2(2)-\frac{123}{2}\ln(3)
       +4\Li_2(1/4)\biggr]
       \nonumber\\*
 &\quad{}-\frac{C_AT_Rn_f}{27}\left(89+3\pi^2\right)\Biggr\}+O(a_s^3).
 \label{eq:gluon-fundamental-magnetic}
\end{align}

The $C_F$-dependent terms in the field and vertex factors cancel in
$c_F$. The coefficient $c_{\parallel\perp}$ has no $n_f$ dependence through
NNLO. The difference $c_{\perp\perp}-c_F$ is
\begin{equation}
 c_{\perp\perp}-c_F
 =a_s^2 C_A^2
       \Bigl[-\frac{2\pi^2}{3}L+\frac{2\pi^2}{3}(\ln(2)-1)
       -\frac{4\pi}{3}\Cl(\pi/3)\Bigr]+O(a_s^3).
 \label{eq:gluon-magnetic-difference}
\end{equation}

For $N_c=3$, the gluon-current coefficients are numerically
\begin{align}
 c_{\parallel\perp}\simeq{}&1-a_s^2(59.2176L+85.8105)+O(a_s^3),
 \label{eq:gluon-parallel-numerical}\\
 c_{\perp\perp}\simeq{}&1+3a_sL+a_s^2\Bigl[(21-n_f)L^2
 -\bigl(25.2176+\tfrac{13}{3}n_f\bigr)L\nonumber\\*
 &\hphantom{1+3a_sL+a_s^2\Bigl[}{}-82.1132-6.58938n_f\Bigr]+O(a_s^3),
 \label{eq:gluon-perp-numerical}\\
 c_F\simeq{}&1+3a_sL+a_s^2\Bigl[(21-n_f)L^2
 +\bigl(34-\tfrac{13}{3}n_f\bigr)L\nonumber\\*
 &\hphantom{1+3a_sL+a_s^2\Bigl[}{}-25.6797-6.58938n_f\Bigr]+O(a_s^3).
 \label{eq:gluon-fundamental-numerical}
\end{align}

\subsubsection{Off-lightcone Wilson-line operators}
The matching coefficients for the bilocal operators follow from
eqs.~\eqref{eq:bilocal-coefficient} and
\eqref{eq:gluon-bilocal-coefficients}. Expanding the products through
$a_s^2$, we obtain for $N_c=3$
\begin{align}
 \mathcal C_\psi\simeq{}&1-a_s(4L+10.7579)\nonumber\\*
 &\hphantom{1}{}+a_s^2\Bigl[\bigl(-14+\tfrac43n_f\bigr)L^2
 +(-123.998+9.39416n_f)L\nonumber\\*
 &\hphantom{1{}+a_s^2\Bigl[}{}-75.234+10.6603n_f\Bigr]+O(a_s^3),
 \label{eq:bilocal-quark-numerical}\\
 \mathcal C_{\parallel\perp}\simeq{}&1-a_s^2(118.435L+171.621)+O(a_s^3),
 \label{eq:bilocal-parallel-numerical}\\
 \mathcal C_{\perp\perp}\simeq{}&1+6a_sL+a_s^2\Bigl[(51-2n_f)L^2
 -\bigl(50.4353+\tfrac{26}{3}n_f\bigr)L\nonumber\\*
 &\hphantom{1+6a_sL+a_s^2\Bigl[}{}-164.226-13.1788n_f\Bigr]+O(a_s^3),
 \label{eq:bilocal-perp-numerical}\\
 \mathcal C_{\parallel\perp,\perp\perp}\simeq{}&1+3a_sL
 +a_s^2\Bigl[(21-n_f)L^2
 -\bigl(84.4353+\tfrac{13}{3}n_f\bigr)L\nonumber\\*
 &\hphantom{1+3a_sL+a_s^2\Bigl[}{}-167.924-6.58938n_f\Bigr]+O(a_s^3).
 \label{eq:bilocal-mixed-numerical}
\end{align}
The two mixed coefficients are equal,
$\mathcal C_{\perp\perp,\parallel\perp}
=\mathcal C_{\parallel\perp,\perp\perp}$.

The following checks support these results:
\begin{itemize}
 \item The gauge parameter cancels in every current coefficient,
 although the separate field and vertex factors are gauge dependent.
 \item Each matching factor is finite after ultraviolet renormalization
 and the infrared subtraction described in section~\ref{sec:procedure}.
 The field and vertex factors also satisfy their renormalization-group
 equations, including the running of the gauge parameter.
 \item Using the same calculation procedure, we have reproduced
 the two-loop matching coefficients for the five flavour-nonsinglet quark
 bilinear operators in the massless limit~\cite{Borgulat:2023bca}.
 \item As an additional check, the field and vertex matching factors for
 the chromoelectric insertion
 $\bar h_v^{\mathrm F}F_{\parallel\perp}^{\mu\nu}h_v^{\mathrm F}$
 on a fundamental Wilson line combine to give unity through NNLO; see
 appendix~\ref{app:fundamental-electric-check}.
\end{itemize}

\subsection{Renormalization-group improvement and numerical analysis}
\label{sec:numerics}
We compare fixed-order and renormalization-group (RG) improved matching
coefficients for $N_c=3$ and $n_f=3$ at $\mu=2\,\mathrm{GeV}$.
We take $\alpha_s(2\,\mathrm{GeV})=0.293$ in the $\MS$ scheme,
obtained from the QCD scale parameter
$\Lambda_{\MS}^{(n_f=3)}=332\,\mathrm{MeV}$~\cite{Petreczky:2022coupling}
using the five-loop beta function~\cite{Baikov:2017fiveLoop}.

\begin{figure}[!t]
 \centering
 \begin{minipage}[t]{0.49\textwidth}
  \centering
  \includegraphics[width=\linewidth]{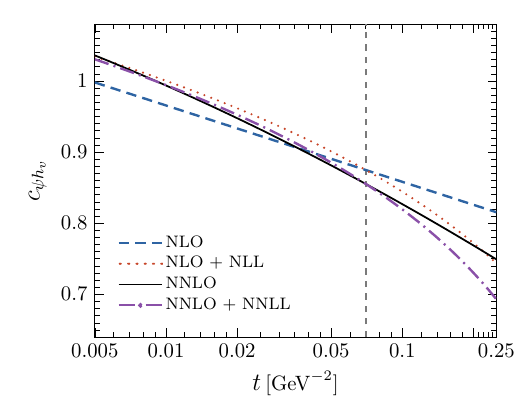}\par
  \raggedright\hspace*{0.12\linewidth}(a)
 \end{minipage}\hfill
 \begin{minipage}[t]{0.49\textwidth}
  \centering
  \includegraphics[width=\linewidth]{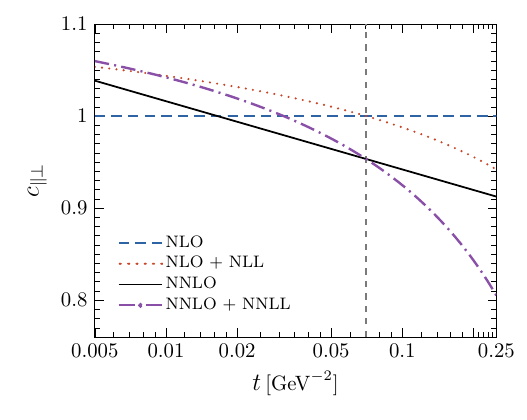}\par
  \raggedright\hspace*{0.12\linewidth}(b)
 \end{minipage}\par\medskip
 \begin{minipage}[t]{0.49\textwidth}
  \centering
  \includegraphics[width=\linewidth]{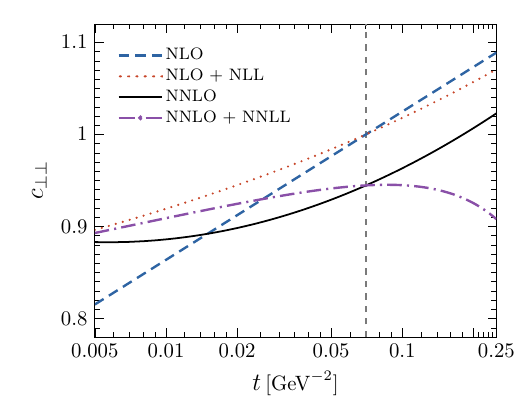}\par
  \raggedright\hspace*{0.12\linewidth}(c)
 \end{minipage}\hfill
 \begin{minipage}[t]{0.49\textwidth}
  \centering
  \includegraphics[width=\linewidth]{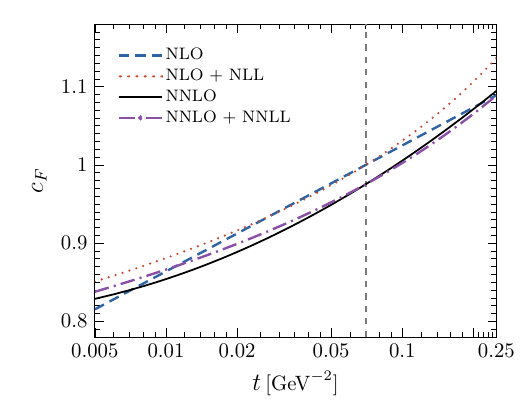}\par
  \raggedright\hspace*{0.12\linewidth}(d)
 \end{minipage}
 \caption{Local-current matching coefficients at $\mu=2\,\mathrm{GeV}$:
 $c_{\psi h_v}$ (a), $c_{\parallel\perp}$ (b),
 $c_{\perp\perp}$ (c) and $c_F$ (d).
 Each panel compares the fixed-order NLO and NNLO results with the
 corresponding RG-improved NLO+NLL and NNLO+NNLL results. The vertical lines mark
 $t\simeq0.07018\,\mathrm{GeV}^{-2}$, where
 $L=\ln(2t\mu^2e^{\gamma_E})=0$ for $\mu=2\,\mathrm{GeV}$.
 The fixed-order NLO result for $c_{\parallel\perp}$ is unity.}
 \label{fig:current-coefficients}
\end{figure}

The matching coefficient $c_J$ compensates the scale dependence of the
corresponding current renormalized in the $\MS$ scheme. Its evolution
from an initial matching
scale $\mu_0$ to the fixed scale $\mu$ is
\begin{equation}
 c_J(t,\mu)=c_J(t,\mu_0)
 \exp\!\left[\int_{a_s(\mu_0)}^{a_s(\mu)}
             \frac{\gamma_J(a_s)}{\beta(a_s)}\,\dd a_s\right].
 \label{eq:local-current-rg}
\end{equation}
We use the anomalous-dimension convention in
eq.~\eqref{eq:current-anomalous-dimension-convention} and define
$\beta(a_s)=\mu\,\dd a_s/\dd\mu$.
The anomalous-dimension and beta-function coefficients are collected in
appendix~\ref{app:renormalization}.

We choose $\mu_0=e^{-\gamma_E/2}/\sqrt{2t}$, so that
$L=\ln(2t\mu_0^2e^{\gamma_E})=0$ in the initial matching coefficient.
At fixed order, we truncate the matching coefficients and their inverses
at next-to-leading order (NLO) and NNLO, retaining terms through $O(a_s)$
and $O(a_s^2)$, respectively.
For RG improvement at next-to-leading-logarithmic (NLL) accuracy,
NLO+NLL combines one-loop matching at $\mu_0$ with two-loop running of
$\alpha_s$ and two-loop current anomalous dimensions. At
next-to-next-to-leading-logarithmic (NNLL) accuracy, NNLO+NNLL combines
two-loop matching with three-loop running and three-loop anomalous
dimensions.
The initial matching coefficients and their inverses are truncated at the
corresponding fixed order, while the RG exponential is kept unexpanded.

We use $0.005\leq t\leq0.25\,\mathrm{GeV}^{-2}$, a range that includes
representative flow times used in recent lattice
calculations~\cite{Brambilla:2026pwave,Bollweg:2026mass}.
The vertical line at $t\simeq0.07018\,\mathrm{GeV}^{-2}$ corresponds to
$\mu_0=\mu$, where $L=\ln(2t\mu^2e^{\gamma_E})=0$ and the
fixed-order and RG-improved results agree at each order.

\begin{figure}[!t]
 \centering
 \begin{minipage}[t]{0.49\textwidth}
  \centering
  \includegraphics[width=\linewidth]{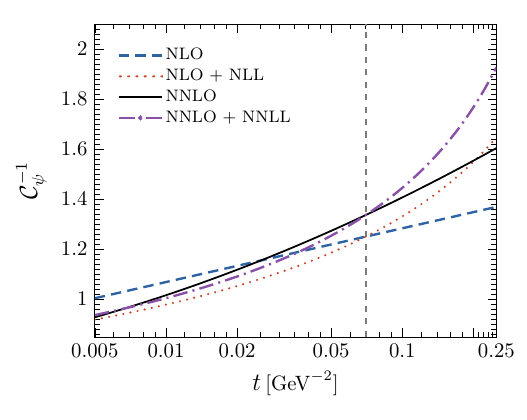}\par
  \raggedright\hspace*{0.12\linewidth}(a)
 \end{minipage}\hfill
 \begin{minipage}[t]{0.49\textwidth}
  \centering
  \includegraphics[width=\linewidth]{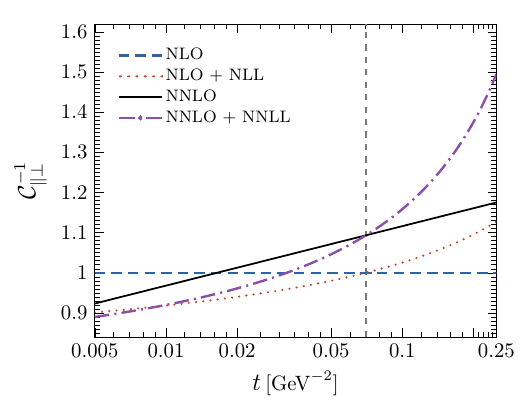}\par
  \raggedright\hspace*{0.12\linewidth}(b)
 \end{minipage}\par\medskip
 \begin{minipage}[t]{0.49\textwidth}
  \centering
  \includegraphics[width=\linewidth]{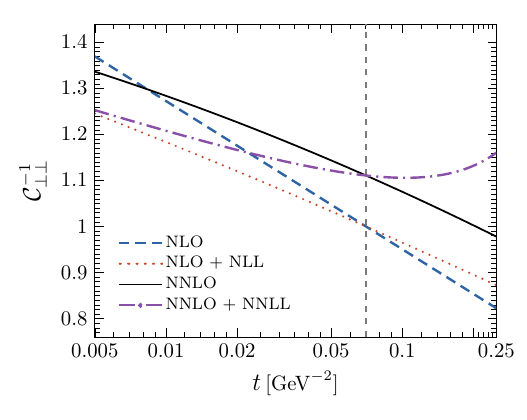}\par
  \raggedright\hspace*{0.12\linewidth}(c)
 \end{minipage}\hfill
 \begin{minipage}[t]{0.49\textwidth}
  \centering
  \includegraphics[width=\linewidth]{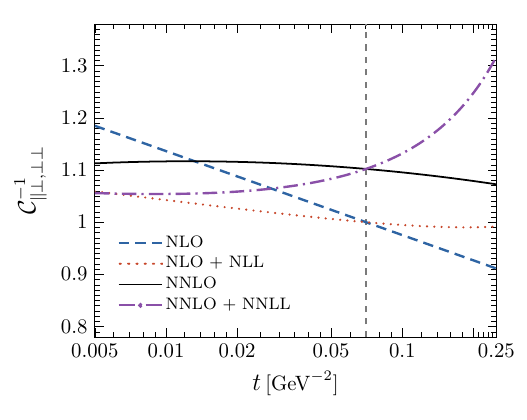}\par
  \raggedright\hspace*{0.12\linewidth}(d)
 \end{minipage}
 \caption{Inverse matching coefficients of the off-lightcone Wilson-line
 operators at $\mu=2\,\mathrm{GeV}$:
 $\mathcal C_\psi^{-1}$ (a),
 $\mathcal C_{\parallel\perp}^{-1}$ (b),
 $\mathcal C_{\perp\perp}^{-1}$ (c) and
 $\mathcal C_{\parallel\perp,\perp\perp}^{-1}$ (d).
 The mixed coefficient satisfies
 $\mathcal C_{\parallel\perp,\perp\perp}
 =\mathcal C_{\perp\perp,\parallel\perp}$.
 Each panel compares the fixed-order NLO and NNLO results with the
 corresponding RG-improved NLO+NLL and NNLO+NNLL results.
 The vertical lines mark
 $t\simeq0.07018\,\mathrm{GeV}^{-2}$, where
 $L=\ln(2t\mu^2e^{\gamma_E})=0$ for $\mu=2\,\mathrm{GeV}$.}
 \label{fig:inverse-coefficients}
\end{figure}

Figures~\ref{fig:current-coefficients} and~\ref{fig:inverse-coefficients}
show the local-current matching coefficients and the inverse matching
coefficients of the off-lightcone Wilson-line operators, respectively.
The inverse coefficients follow from the current products in
eqs.~\eqref{eq:bilocal-coefficient} and \eqref{eq:gluon-bilocal-coefficients}.
The RG evolution of these products follows from the sums of the
corresponding current anomalous dimensions.

\Needspace{6\baselineskip}
We focus on the change from NLO+NLL to NNLO+NNLL at fixed $t$.
We quote the absolute difference between these results, divided by the
magnitude of the full NLO+NLL result, as a percentage. For fixed-order
comparisons, the absolute difference between the NNLO and NLO results is
instead divided by the magnitude of the full NLO result.
For $t\lesssim0.1\,\mathrm{GeV}^{-2}$
within the plotted range, the changes reach at most about $7\%$ for the
local currents and $15\%$ for the inverse matching coefficients.
We compare the results over finite ranges of $t$, since agreement at an
isolated crossing does not establish perturbative convergence.

For $c_F$, the relative changes from NLO+NLL to NNLO+NNLL remain below
$4.3\%$ throughout the plotted range. The corresponding fixed-order changes
from NLO to NNLO remain below $3\%$.

For the quark-current coefficient, the RG-improved changes remain below
$3\%$ for $t\lesssim0.1\,\mathrm{GeV}^{-2}$ within the plotted range,
rising to about $7\%$ at $t=0.25\,\mathrm{GeV}^{-2}$. The corresponding
inverse coefficient $\mathcal C_\psi^{-1}$ shows larger changes, up to
about $9\%$ and $17\%$, respectively.

For the adjoint-current coefficients and the inverse gluon matching coefficients,
the relative changes increase towards the upper end of the plotted flow-time
range. Below $t\simeq0.04\,\mathrm{GeV}^{-2}$ within this range, the
RG-improved changes are at most about $4\%$ for the currents and $7\%$ for
the inverse coefficients. Over $t\simeq0.04$--$0.1\,\mathrm{GeV}^{-2}$,
they grow to about $3$--$7\%$ and $5$--$15\%$, respectively. At
$t=0.25\,\mathrm{GeV}^{-2}$, the relative changes are about $15\%$ for both
adjoint-current coefficients and about $33\%$ for all three inverse matching
coefficients in the gluon sector.

\section{Summary}\label{sec:discussion}
We have calculated, in the small-flow-time limit, the matching coefficients required to relate mass-subtracted flowed
off-lightcone Wilson-line operators to their $\MS$-renormalized
counterparts through NNLO in QCD. Using the auxiliary-field formalism, we reduce
the calculation to four local currents: the quark current $J_q$, the
two projected field-strength currents $J_{\parallel\perp}$ and
$J_{\perp\perp}$ with an adjoint auxiliary field, and the
chromomagnetic insertion $J_F$ on a fundamental Wilson line. The flowed
quark fields are defined in the ringed scheme.

After mass subtraction, the matching coefficients for quark and gluon
Wilson-line operators are products of the corresponding local-current
coefficients when the separation is large compared with the flow radius
$\sqrt{8t}$. Combining the matching coefficients with three-loop
anomalous dimensions and the beta function, we obtain their RG improvement
at NNLL accuracy. In the numerical examples, the changes from NLO+NLL to
NNLO+NNLL remain below $7\%$ for the quark-current coefficient and below
$4.3\%$ for $c_F$ throughout the plotted range. The relative changes for the
adjoint-current coefficients and the inverse gluon matching coefficients
increase towards the upper end of that range.

These results can be applied to lattice calculations of unpolarized and
polarized quark and gluon quasi-parton distributions and meson
quasi-distribution amplitudes. Products of quark-current coefficients
also enter the matching of three-quark operators used for baryon
quasi-distribution amplitudes. Other
applications include gluonic correlators for heavy-quark diffusion and
quarkonium transport, quarkonium decay and production, the static force,
spin-dependent heavy-quark potentials and hybrid--quarkonium mixing.

\acknowledgments
This work is supported by the National Natural Science Foundation of China
under Grant Nos. 12675114 and 12635005.

The author used large language models (LLMs) to assist with computational
code development and debugging, selected algebraic manipulations and
language editing. The author is responsible
for the scientific content and conclusions.

\appendix

\section{Renormalization constants and anomalous dimensions}
\label{app:renormalization}
We use the $\MS$ scheme and covariant $R_\xi$ gauge, with the following
relations between bare and renormalized quantities:
\begin{alignat}{2}
 \psi_0&=\sqrt{Z_\psi}\psi,\qquad
 &h_{v,0}&=\sqrt{Z_{h_v}}h_v,\nonumber\\
 J_q^{\rm B}&=Z_{J_q}J_q,\qquad
 &\chi_0&=\sqrt{Z_\chi}\chi.
 \label{eq:app-z-conventions}
\end{alignat}
\subsection{Parameters and fields}
We use $a_s=\alpha_s(\mu)/(4\pi)$ and define the beta function by
$\beta(a_s)=\dd a_s/\dd\ln(\mu)
=-2\beta_0a_s^2-2\beta_1a_s^3-2\beta_2a_s^4+O(a_s^5)$.

The standard renormalization relations for the coupling and gauge parameter are
\begin{alignat}{2}
 g_0&=\widetilde\mu^\epsilon Z_g g,\qquad&
 Z_g&=1-\frac{a_s\beta_0}{2\epsilon}
 +a_s^2\left(\frac{3\beta_0^2}{8\epsilon^2}-\frac{\beta_1}{4\epsilon}\right)+O(a_s^3),\nonumber\\
 \xi_0&=Z_3\xi,&
 \beta_0&=\frac{11}{3}C_A-\frac43T_Rn_f,\nonumber\\
 &&\beta_1&=\frac{34}{3}C_A^2
                  -\left(\frac{20}{3}C_A+4C_F\right)T_Rn_f.
 \label{eq:coupling-renormalization}
\end{alignat}
Here $n_f$ denotes the number of quark flavours. The three-loop
coefficient is~\cite{Larin:1993tp,Tarasov:1980beta}
\begin{align}
 \beta_2=&\frac{2857}{54}C_A^3
 +\left(2C_F^2-\frac{205}{9}C_FC_A-\frac{1415}{27}C_A^2\right)T_Rn_f
 \nonumber\\*
 &{}+\left(\frac{44}{9}C_F+\frac{158}{27}C_A\right)(T_Rn_f)^2.
 \label{eq:app-beta-three}
\end{align}

\paragraph{Gauge field.}
The gluon-field renormalization constant is~\cite{Larin:1993tp}
\begin{align}
 Z_3={}&1+\frac{a_s}{\epsilon}
       \left[C_A\left(\frac{13}{6}-\frac\xi2\right)-\frac43T_Rn_f\right]
 \nonumber\\*
 &\hphantom{1}{}+\frac{a_s^2}{\epsilon^2}\left[
 C_A^2\left(-\frac{13}{8}-\frac{17\xi}{24}+\frac{\xi^2}{4}\right)
 +C_AT_Rn_f\left(1+\frac{2\xi}{3}\right)\right]\nonumber\\*
 &\hphantom{1}{}+\frac{a_s^2}{\epsilon}\left[
 C_A^2\left(\frac{59}{16}-\frac{11\xi}{16}-\frac{\xi^2}{8}\right)
 -2C_FT_Rn_f-\frac52C_AT_Rn_f\right]+O(a_s^3).
 \label{eq:app-z3}
\end{align}
With the gauge coupling absorbed into the gauge field,
$A_0^\mu=Z_g\sqrt{Z_3}A^\mu$.
The gauge parameter used in ref.~\cite{Larin:1993tp} corresponds to $1-\xi$.

\paragraph{Unflowed light-quark field.}
The quark-field renormalization constant is~\cite{QCDRenorm,Larin:1993tp}
\begin{align}
 Z_\psi={}&1-\frac{a_sC_F\xi}{\epsilon}
 +\frac{a_s^2}{\epsilon^2}
       \left[\frac12C_F^2\xi^2+\frac14C_FC_A\xi(3+\xi)\right]
 \nonumber\\*
 &\hphantom{1}{}+\frac{a_s^2}{\epsilon}
       \left[\frac34C_F^2-C_FC_A\left(\frac{25}{8}+\xi+\frac{\xi^2}{8}\right)
                              +C_FT_Rn_f\right]+O(a_s^3).
 \label{eq:app-z2}
\end{align}
\paragraph{Flowed light-quark field and conversion to the ringed scheme.}
For the flowed field, our $Z_\chi$ is the inverse of the $\MS$
renormalization constant $Z_\chi^{\MS}$ in ref.~\cite{Borgulat:2023bca}.
The $\MS$ factor $Z_\chi$ is independent of both QCD and flow gauge parameters.
This follows to all orders from the finiteness of gauge-invariant flowed
quark bilinear operators after field renormalization~\cite{Hieda:2016renormalizability}
and the uniqueness of their $\MS$ pole subtraction.
The two-loop result was obtained in ref.~\cite{Harlander:2018zpi} and is
also given in ref.~\cite{Artz:2019bpr}:
\begin{align}
 Z_\chi={}&1-\frac{3a_sC_F}{\epsilon}
 +\frac{a_s^2}{\epsilon^2}\left(\frac92C_F^2+\frac32\beta_0C_F\right)
 \nonumber\\*
 &{}-\frac{a_s^2C_F}{4\epsilon}\left[
 C_A\left(\frac{223}{3}-16\ln(2)\right)
 -C_F\left(3+16\ln(2)\right)-\frac{44}{3}T_Rn_f\right]+O(a_s^3).
 \label{eq:app-zchi}
\end{align}
The conversion to the ringed field is finite:
$\mathring\chi=\sqrt{\mathring Z_\chi Z_\chi}\,\chi$, where
\cite{Borgulat:2023bca,Artz:2019bpr}
\begin{align}
 \mathring Z_\chi Z_\chi={}&1+a_sC_F\bigl[3L-\ln(432)\bigr]\nonumber\\*
 &\hphantom{1}{}+a_s^2\Biggl\{\frac{C_F(3C_F+\beta_0)}2
 L\bigl(3L-2\ln(432)\bigr)\nonumber\\*
 &\quad{}+\frac{C_FL}{2}\left[
 C_A\left(\frac{223}{3}-16\ln(2)\right)
 -C_F\left(3+16\ln(2)\right)-\frac{44}{3}T_Rn_f\right]\nonumber\\*
 &\quad{}+C_{\chi,2}\Biggr\}+O(a_s^3).
 \label{eq:app-ringed}
\end{align}
The constant $C_{\chi,2}$ is given numerically in
eq.~\eqref{eq:cchi-numerical}.

\paragraph{Auxiliary field.}
For representation $R$, let $C_R$ denote its quadratic Casimir, with
$C_R=C_F$ or $C_A$ for the fundamental or adjoint representation.
The result for the fundamental representation in refs.~\cite{BG,Chetyrkin:2003vi}
extends to representation $R$ through two loops:
\begin{align}
Z_{h_v}^{R}={}&1+\frac{a_sC_R(3-\xi)}\epsilon
 +\frac{a_s^2}{\epsilon^2}\left[
 \frac12C_R^2(3-\xi)^2
 +C_RC_A\left(-\frac{11}{2}+\frac{3\xi}{4}+\frac{\xi^2}{4}\right)
 +2C_RT_Rn_f\right]\nonumber\\*
 &\hphantom{1}{}+\frac{a_s^2}{\epsilon}\left[
 C_RC_A\left(\frac{179}{24}-\xi-\frac{\xi^2}{8}\right)
                         -\frac83C_RT_Rn_f\right]+O(a_s^3).
 \label{eq:app-zh-representation}
\end{align}

\subsection{Current operators}
The current renormalization constants are defined by $J_X^{\rm B}=Z_XJ_X^{\MS}$
in physical matrix elements.

\paragraph{Quark current.}
For $J_q=\bar\psi h_v^{\mathrm F}$, the renormalization constant
is~\cite{BG,Ji:1991heavyLight}
\begin{align}
Z_{J_q}={}&1+\frac{3a_sC_F}{2\epsilon}
 +\frac{a_s^2}{\epsilon^2}
       \left(\frac98C_F^2-\frac{11}{4}C_FC_A+C_FT_Rn_f\right)
 \nonumber\\*
 &\hphantom{1}{}+\frac{a_s^2}{\epsilon}\left[
 C_F^2\left(-\frac58+\frac{2\pi^2}{3}\right)
 +C_FC_A\left(\frac{49}{24}-\frac{\pi^2}{6}\right)
 -\frac56C_FT_Rn_f\right]+O(a_s^3).
 \label{eq:app-zh-zj}
\end{align}

\paragraph{Gluon currents.}
The two-loop renormalization constants for $J_{\parallel\perp}$ and
$J_{\perp\perp}$ were obtained in ref.~\cite{Braun:2020ymy}; the
corresponding expressions are also given in ref.~\cite{Braun:2021gluon}.
The result for $J_F$ is taken from ref.~\cite{Amoros:1997mag}:
\begin{align}
 Z_{\parallel\perp}
 &=1+\frac{a_s^2\pi^2C_A^2}{3\epsilon}+O(a_s^3),
 \label{eq:app-z-current-parallel}\\[3pt]
 Z_{\perp\perp}
 &=1-\frac{a_sC_A}{\epsilon}
 +\frac{a_s^2}{\epsilon^2}\left(\frac73C_A^2-\frac23C_AT_Rn_f\right)
 \nonumber\\*
 &\hphantom{{}=1}{}+\frac{a_s^2}{\epsilon}\left[
 C_A^2\left(-\frac{17}{9}+\frac{\pi^2}{3}\right)
 +\frac{13}{9}C_AT_Rn_f\right]+O(a_s^3),
 \label{eq:app-z-current-perp}\\[3pt]
 Z_F
 &=1-\frac{a_sC_A}{\epsilon}
 +\frac{a_s^2}{\epsilon^2}\left(\frac73C_A^2-\frac23C_AT_Rn_f\right)
 \nonumber\\*
 &\hphantom{{}=1}{}+\frac{a_s^2}{\epsilon}\left(-\frac{17}{9}C_A^2
 +\frac{13}{9}C_AT_Rn_f\right)+O(a_s^3).
 \label{eq:app-gluon-current-z}
\end{align}

\subsection{Amputated current vertices}
The renormalization constants $Z_X^{\rm vert}$ for the amputated current
vertices are obtained by dividing the current renormalization constants
by the corresponding external-field renormalization factors. An unflowed
bare vertex equals $Z_X^{\rm vert}$ times its renormalized counterpart.
The subscript E denotes the chromoelectric insertion
$\bar h_v^{\mathrm F}F_{\parallel\perp}^{\mu\nu}h_v^{\mathrm F}$.
With the gauge coupling absorbed into the gauge field, we have
\begin{align}
 Z_{\psi h_v}^{\rm vert}
 &=\frac{Z_{J_q}}{\sqrt{Z_\psi Z_{h_v}^{\mathrm F}}},\qquad
 Z_r^{\rm vert}=\frac{Z_r}{Z_g\sqrt{Z_3Z_{h_v}^{\mathrm A}}},
 \quad r=\parallel\perp,\perp\perp,\nonumber\\
 Z_F^{\rm vert}&=\frac{Z_F}{Z_g\sqrt{Z_3}Z_{h_v}^{\mathrm F}},\qquad
 Z_E^{\rm vert}=\frac{1}{Z_g\sqrt{Z_3}Z_{h_v}^{\mathrm F}}+O(a_s^3).
 \label{eq:app-vertex-z-ratios}
\end{align}
The convention for $Z_{\psi h_v}^{\rm vert}$ follows
refs.~\cite{BG,Chetyrkin:2003vi}.
The relation for $Z_E^{\rm vert}$~\cite{Amoros:1997velocity} holds
through NNLO for the physical projection, after separating contributions
from equation-of-motion operators. The flowed-quark renormalization constant
and finite ringed normalization enter only the field conversion.
Expanding these ratios gives
\begin{align}
Z_{\psi h_v}^{\rm vert}={}&1+\frac{a_s C_F\xi}{\epsilon}+\frac{a_s^2}{\epsilon^2}\left[
 \frac{C_F^2\xi^2}{2}-\frac{C_FC_A\xi(3+\xi)}4\right]\nonumber\\*
 &\hphantom{1}{}+\frac{a_s^2}{\epsilon}\left[
 C_F^2\left(-1+\frac{2\pi^2}{3}\right)
 +C_FC_A\left(-\frac18-\frac{\pi^2}{6}+\xi+\frac{\xi^2}{8}\right)
 \right]+O(a_s^3),
 \label{eq:app-quark-vertex-z}\\[3pt]
Z_{\parallel\perp}^{\rm vert}={}&1+\frac{3a_sC_A(\xi-1)}{4\epsilon}+\frac{a_s^2}{\epsilon^2}\left[
 \frac{C_A^2}{32}(53-36\xi+3\xi^2)-\frac12C_AT_Rn_f\right]\nonumber\\*
 &\hphantom{1}{}+\frac{a_s^2}{\epsilon}\left[
 \frac{C_A^2}{96}(-263+81\xi+12\xi^2+32\pi^2)
 +\frac{11}{12}C_AT_Rn_f\right]+O(a_s^3),
 \label{eq:app-adjoint-electric-vertex-z}\\[3pt]
Z_{\perp\perp}^{\rm vert}={}&1+\frac{a_sC_A(3\xi-7)}{4\epsilon}+\frac{a_s^2}{\epsilon^2}\left[
 \frac{C_A^2}{96}(455-180\xi+9\xi^2)-\frac76C_AT_Rn_f\right]\nonumber\\*
 &\hphantom{1}{}+\frac{a_s^2}{\epsilon}\left[
 \frac{C_A^2}{288}(-1333+243\xi+36\xi^2+96\pi^2)
 +\frac{85}{36}C_AT_Rn_f\right]+O(a_s^3),
 \label{eq:app-adjoint-magnetic-vertex-z}\\[3pt]
Z_F^{\rm vert}={}&1+\frac{a_s}{\epsilon}
 \biggl[\frac{C_A(\xi-1)}4+C_F(\xi-3)\biggr]\nonumber\\*
 &\hphantom{1}{}+\frac{a_s^2}{\epsilon^2}\biggl[
 \frac{C_F^2(\xi-3)^2}{2}+\frac{C_AC_F(25-7\xi)}4
 +\frac{C_A^2}{96}(47-24\xi-3\xi^2)\nonumber\\*
 &\hphantom{1}\qquad{}-2C_FT_Rn_f-\frac16C_AT_Rn_f\biggr]\nonumber\\*
 &\hphantom{1}{}+\frac{a_s^2}{\epsilon}\biggl[
 C_AC_F\left(-\frac{179}{24}+\xi+\frac{\xi^2}{8}\right)
 +\frac{C_A^2}{288}(-259+99\xi+18\xi^2)\nonumber\\*
 &\hphantom{1}\qquad{}+\frac83C_FT_Rn_f+\frac{37}{36}C_AT_Rn_f\biggr]+O(a_s^3),
 \label{eq:app-fundamental-magnetic-vertex-z}\\[3pt]
Z_E^{\rm vert}={}&1+\frac{a_s}{\epsilon}
 \biggl[\frac{C_A(3+\xi)}4+C_F(\xi-3)\biggr]\nonumber\\*
 &\hphantom{1}{}+\frac{a_s^2}{\epsilon^2}\biggl[
 \frac{C_F^2(\xi-3)^2}{2}+\frac{C_AC_F(13-3\xi)}4
 -\frac{C_A^2}{32}(35+\xi^2)-2C_FT_Rn_f+\frac12C_AT_Rn_f\biggr]\nonumber\\*
 &\hphantom{1}{}+\frac{a_s^2}{\epsilon}\biggl[
 C_AC_F\left(-\frac{179}{24}+\xi+\frac{\xi^2}{8}\right)
 +\frac{C_A^2}{96}(95+33\xi+6\xi^2)\nonumber\\*
 &\hphantom{1}\qquad{}+\frac83C_FT_Rn_f-\frac5{12}C_AT_Rn_f\biggr]+O(a_s^3).
 \label{eq:app-fundamental-electric-vertex-z}
\end{align}

\subsection{Current anomalous dimensions}
We define $\gamma_J=\dd\ln(Z_J)/\dd\ln(\mu)$ and expand
$\gamma_J=a_s\gamma_{J,0}+a_s^2\gamma_{J,1}+a_s^3\gamma_{J,2}+\cdots$.
The one- and two-loop terms
are~\cite{Braun:2020ymy,BG,Ji:1991heavyLight,Amoros:1997mag}
\begin{align}
\gamma_{J_q}&={}-3a_sC_F\nonumber\\*
 &\hphantom{{}=}{}+a_s^2C_F\left[
 C_F\left(\frac52-\frac{8\pi^2}{3}\right)
 +C_A\left(\frac{2\pi^2}{3}-\frac{49}{6}\right)
 +\frac{10}{3}T_Rn_f\right]+O(a_s^3),
 \label{eq:current-anomalous-dimension}\\[3pt]
 \gamma_{\parallel\perp}
 &=-\frac43\pi^2C_A^2a_s^2+O(a_s^3),\nonumber\\
 \gamma_{\perp\perp}
 &=2C_Aa_s+a_s^2\left[C_A^2\left(\frac{68}{9}-\frac{4\pi^2}{3}\right)
                  -\frac{52}{9}C_AT_Rn_f\right]+O(a_s^3),
 \label{eq:app-adjoint-gamma}\\[3pt]
 \gamma_F
 &=2C_Aa_s+a_s^2\left(\frac{68}{9}C_A^2-\frac{52}{9}C_AT_Rn_f\right)
                  +O(a_s^3).
 \label{eq:app-fundamental-gamma}
\end{align}

For $\mathrm{SU}(N_c)$ with $T_R=1/2$, the three-loop coefficients are
\begin{align}
 \gamma_{J_q,2}&={}C_F^3\left(-\frac{37}{2}+\frac{32\pi^2}{3}
 -36\zeta(3)-\frac{8\pi^4}{9}\right)\nonumber\\*
 &\hphantom{{}={}}{}+C_F^2C_A\left(-\frac{655}{36}-\frac{592\pi^2}{27}
 +\frac{142}{3}\zeta(3)-\frac{8\pi^4}{45}\right)\nonumber\\*
 &\hphantom{{}={}}{}+C_FC_A^2\left(\frac{1451}{108}+\frac{130\pi^2}{27}
 -\frac{22}{3}\zeta(3)-\frac{4\pi^4}{15}\right)\nonumber\\
 &\hphantom{{}={}}{}+C_F^2T_Rn_f\left(\frac{470}{9}+\frac{224\pi^2}{27}
 -\frac{176}{3}\zeta(3)\right)\nonumber\\*
 &\hphantom{{}={}}{}+C_FC_AT_Rn_f\left(-\frac{512}{27}-\frac{56\pi^2}{27}
 +\frac{152}{3}\zeta(3)\right)+\frac{140}{27}C_F(T_Rn_f)^2,
 \label{eq:fixed-mu-gamma-three}\\[3pt]
 \gamma_{\parallel\perp,2}&={}C_A^3\left(-\frac{16\pi^2}{27}
 -\frac{4\pi^4}{3}\right)
 +C_A^2T_Rn_f\left(-\frac{88\pi^2}{27}+\frac{8\pi^4}{9}\right),
 \label{eq:app-adjoint-electric-gamma-three}\\[3pt]
 \gamma_{\perp\perp,2}&={}C_A^3\left(\frac{899}{27}+\frac{20\pi^2}{27}
 -\frac{4\pi^4}{3}+8\zeta(3)\right)\nonumber\\*
 &\hphantom{{}={}}{}+C_A^2T_Rn_f\left(-\frac{520}{27}-\frac{64\pi^2}{27}
 +\frac{8\pi^4}{9}-32\zeta(3)\right)\nonumber\\*
 &\hphantom{{}={}}{}+C_AC_FT_Rn_f\left(-\frac{98}{3}+32\zeta(3)\right)
 -\frac{16}{9}C_AT_R^2n_f^2,
 \label{eq:app-adjoint-magnetic-gamma-three}\\[3pt]
 \gamma_{F,2}&={}C_A^3\left(\frac{899}{27}+8\zeta(3)\right)
 +C_A^2T_Rn_f\left(-\frac{520}{27}-32\zeta(3)\right)\nonumber\\*
 &\hphantom{{}={}}{}+C_AC_FT_Rn_f\left(-\frac{98}{3}+32\zeta(3)\right)
 -\frac{16}{9}C_AT_R^2n_f^2\nonumber\\*
 &\hphantom{{}={}}{}+\frac{4\pi^2}{3}N_c(N_c^2+6)
 -\frac{8\pi^2n_f}{9}\frac{N_c^4-6N_c^2+18}{N_c^2}.
 \label{eq:app-fundamental-magnetic-gamma-three}
\end{align}
The three-loop coefficient for the quark current was obtained in
ref.~\cite{Chetyrkin:2003vi}; see also ref.~\cite{Grozin:2023dlk}.
The three-loop coefficients for the adjoint gluon currents are taken from
ref.~\cite{Braun:2021gluon}, and that for the chromomagnetic insertion
from ref.~\cite{Grozin:2007mag}.

\section{Matching coefficients for fields and vertices}\label{app:matching}

\subsection{Field matching coefficients}\label{app:field-matching}
\label{app:gluon-fields}
\paragraph{$\MS$-renormalized flowed light-quark field.}
For $\chi(t)=\zeta_\psi\psi^{\MS}+O(t)$, the matching coefficient is
\begin{align}
 \zeta_\psi={}&1+\frac{a_sC_F}{2}\bigl[(\xi-3)L+\xi-1\bigr]\nonumber\\*
 &\hphantom{1}{}+a_s^2\Biggl\{L^2\biggl[
 \frac{C_F^2}{8}(\xi-3)^2
 +\frac{C_FC_A}{8}\bigl[\xi(\xi+3)-22\bigr]
 +C_FT_Rn_f\biggr]\nonumber\\
 &\quad{}+L\biggl[C_F^2\left(\frac{(\xi-3)(\xi-1)}4+4\ln(2)\right)
 +\frac{C_FC_A}{24}\bigl[-415+42\xi+9\xi^2+96\ln(2)\bigr]
 \nonumber\\*
 &\qquad{}+\frac{10}{3}C_FT_Rn_f\biggr]
 +C_F^2\biggl[\frac{(\xi-1)^2}{8}
 +4\ln(2)+2\ln^2(2)
 +3\Li_2(1/4)-\frac{\pi^2}{6}\biggr]\nonumber\\
 &\quad{}+C_FC_A\biggl[\frac{\pi^2(\xi^2-21)}{96}
 +\frac{-454+51\xi+9\xi^2}{24}
 +\frac52\xi\ln(2)+2\ln^2(2)-26\ln(2)\nonumber\\*
 &\hphantom{1}\qquad {}+\left(27-\frac94\xi\right)\ln(3)
 +\left(\frac{51}{4}-\frac32\xi\right)\Li_2(1/4)\biggr]
 +C_FT_Rn_f\biggl[\frac{10}{3}
                         +\frac{\pi^2}{6}\biggr]\Biggr\}+O(a_s^3).
 \label{eq:res-ms-light}
\end{align}

\Needspace{6\baselineskip}
\paragraph{Ringed flowed light-quark field.}
Using the finite conversion in eq.~\eqref{eq:app-ringed}, we obtain
\begin{align}
 \mathring\zeta_\psi={}&\sqrt{\mathring Z_\chi Z_\chi}\,\zeta_\psi\nonumber\\*
 {}={}&1+\frac{a_sC_F}{2}
 \bigl[\xi L+\xi-1-\ln(432)\bigr]\nonumber\\*
 &\hphantom{1}{}+a_s^2\Biggl\{L^2\left[
 \frac{C_F^2\xi^2}{8}+\frac{C_FC_A\xi(\xi+3)}8\right]\nonumber\\
 &\quad{}+L\biggl[\frac{C_F^2}{4}\bigl[\xi(\xi-1-\ln(432))-3\bigr]
 +\frac{C_FC_A}{24}\bigl[31+42\xi+9\xi^2-44\ln(432)\bigr]
 \nonumber\\*
 &\qquad{}+\frac{C_FT_Rn_f}{3}\bigl[-1+2\ln(432)\bigr]\biggr]\nonumber\\
 &\quad{}+C_F^2\biggl[\frac{(\xi-1)^2}{8}
 +\frac{1-\xi}{4}\ln(432)-\frac18\ln^2(432)
 +4\ln(2)+2\ln^2(2)\nonumber\\*
 &\qquad{}+3\Li_2(1/4)-\frac{\pi^2}{6}\biggr]
 +C_FC_A\biggl[\frac{\pi^2(\xi^2-21)}{96}
 +\frac{-454+51\xi+9\xi^2}{24}\nonumber\\*
 &\qquad{}+\frac52\xi\ln(2)+2\ln^2(2)-26\ln(2)
 +\left(27-\frac94\xi\right)\ln(3)\nonumber\\*
 &\qquad{}+\left(\frac{51}{4}-\frac32\xi\right)\Li_2(1/4)\biggr]
 +\frac{C_FT_Rn_f}{6}\bigl(20+\pi^2\bigr)
 +\frac12C_{\chi,2}\Biggr\}+O(a_s^3).
 \label{eq:res-q2}
\end{align}

\paragraph{Auxiliary field.}
For $h_v=\zeta_{h_v}h_v^{\MS}+O(t)$, the matching coefficient in a general
representation $R$ is
\begin{align}
 \zeta_{h_v}={}&1+\frac{a_sC_R}{2}\bigl[(\xi-3)L+\xi-1\bigr]\nonumber\\*
 &\hphantom{1}{}+a_s^2\Biggl\{L^2\biggl[
 \frac{C_R^2}{8}(\xi-3)^2
 +\frac{C_AC_R}{8}\bigl[\xi(\xi+3)-22\bigr]
 +C_RT_Rn_f\biggr]\nonumber\\
 &\quad{}+L\biggl[\frac{C_R^2}{4}(\xi-3)(\xi-1)
 +\frac{C_AC_R}{24}\bigl[-223+42\xi+9\xi^2\bigr]
 +\frac{10}{3}C_RT_Rn_f\biggr]\nonumber\\
 &\quad{}+\frac{C_R^2}{8}(\xi-1)^2
 +\frac{C_AC_R}{96}\biggl\{
       -\pi(96+\pi)+204\xi+(36+\pi^2)\xi^2
       -2152\nonumber\\*
 &\qquad{}+48(5\xi-106)\ln(2)-216(\xi-24)\ln(3)
 -144(\xi-3)\Li_2(1/4)\nonumber\\*
 &\qquad{}-72\Li_3(1/4)-96\zeta(3)\biggr\}
 +\frac{C_{R}T_R n_f}{6}
       \bigl(20+\pi^2\bigr)\Biggr\}+O(a_s^3).
 \label{eq:res-h2}
\end{align}
Here $C_R=C_F$ for the fundamental representation and $C_R=C_A$ for the
adjoint representation.

\Needspace{6\baselineskip}
\paragraph{Gluon field.}
For $B^\mu(t)=\zeta_A A^{\mu,\MS}+O(t)$, the matching coefficient is
\begin{align}
 \zeta_A={}&1+\frac{a_sC_A}{4}\bigl[(3+\xi)L+1+\xi\bigr]\nonumber\\*
 &\hphantom{1}{}+a_s^2\Biggl\{L^2\left[
 \frac{C_A^2}{32}\bigl[53+3\xi(4+\xi)\bigr]
 -\frac{C_AT_Rn_f}{2}\right]\nonumber\\
 &\quad{}+L\left[\frac{C_A^2}{48}\bigl[148+3\xi(21+5\xi)\bigr]
 -\frac76C_AT_Rn_f\right]\nonumber\\
 &\quad{}+\frac{C_A^2}{384}\biggl\{
       \pi^2(-1+\xi+2\xi^2)
       +8\bigl[265+843\ln(2)-810\ln(3)+27\Li_2(1/4)\bigr]\nonumber\\*
 &\qquad{}+6\xi\bigl[42+13\xi+8(11-\xi)\ln(2)+12(\xi-3)\ln(3)
             -(15+\xi)\Li_2(1/4)\bigr]\biggr\}\nonumber\\
 &\quad{}-\frac{C_A T_R n_f}{24}
       \bigl(25+2\pi^2\bigr)\Biggr\}+O(a_s^3).
 \label{eq:gluon-connection-two-loop}
\end{align}

\subsection{Vertex matching coefficients}\label{app:vertex-matching}
\label{app:gluon-vertices}\label{app:fundamental-electric-check}
\paragraph{Quark current.}
For the quark current $J_q$ in eq.~\eqref{eq:gluon-currents}, the vertex
matching coefficient defined by eq.~\eqref{eq:three-factor} is
\begin{align}
 \zeta_{\psi h_v}={}&1-a_s\bigl[LC_F\xi+C_F\xi\bigr]
 +a_s^2\Biggl\{L^2\left[\frac{C_F^2\xi^2}{2}
 -\frac{C_FC_A\xi(\xi+3)}4\right]\nonumber\\
 &\quad{}+L\left[C_F^2\left(\xi^2+2-\frac{4\pi^2}{3}\right)
 +C_FC_A\left(\frac14+\frac{\pi^2}{3}-\frac{\xi(14+3\xi)}4\right)\right]\nonumber\\
 &\quad{}+C_F^2\biggl[\frac{\xi^2}{2}+\frac32+6\ln(2)-3\ln(3)
 +\frac{\pi^2}{6}\bigl[-7+8\ln(2)\bigr]\nonumber\\*
 &\qquad{}-4\pi\operatorname{Cl}_2(\pi/3)
 +\frac52\Li_2(1/4)-\Li_3(1/4)-\frac43\zeta(3)\biggr]\nonumber\\
 &\quad{}+C_FC_A\biggl[\frac32-6\ln(2)+3\ln(3)
 +\frac{\pi^2}{48}\bigl[19-\xi^2-16\ln(2)\bigr]\nonumber\\*
 &\qquad{}+\pi\operatorname{Cl}_2(\pi/3)
 -\frac14\Li_2(1/4)+\frac14\Li_3(1/4)
 +\frac13\zeta(3)\nonumber\\*
 &\qquad{}-\frac\xi4\bigl[17+3\xi+20\ln(2)-18\ln(3)
 -12\Li_2(1/4)\bigr]\biggr]\Biggr\}+O(a_s^3).
 \label{eq:res-v2}
\end{align}
\paragraph{Adjoint gluon currents.}
The vertex factors defined by eqs.~\eqref{eq:gluon-factorization-parallel}
and~\eqref{eq:gluon-factorization-perp} are
\begin{align}
 \zeta_{\parallel\perp}={}&1+\frac{a_sC_A}{4}\bigl[L(3-3\xi)+1-3\xi\bigr]
 +a_s^2\Biggl\{L^2\left[
 \frac{C_A^2}{32}\bigl(53-36\xi+3\xi^2\bigr)-\frac12C_AT_Rn_f\right]\nonumber\\*
 &\quad{}+L\left[C_A^2\left(\frac{316-171\xi-3\xi^2}{48}
 -\frac{2\pi^2}{3}\right)-\frac{13}{6}C_AT_Rn_f\right]\nonumber\\*
 &\quad{}+\frac{C_A^2}{1152}\biggl\{
 -384\pi\bigl[-11+4\sqrt3+4\Cl(\pi/3)\bigr]
 -\pi^2\bigl[1265+3\xi(1+6\xi)-768\ln(2)\bigr]\nonumber\\*
 &\qquad{}+18\xi\Bigl[-186-17\xi+8(\xi-31)\ln(2)
 -12(\xi-15)\ln(3)+(111+\xi)\Li_2(1/4)\Bigr]\nonumber\\*
 &\qquad{}+24\Bigl[814+1701\ln(2)-1782\ln(3)-243\Li_2(1/4)
 +36\Li_3(1/4)+48\zeta(3)\Bigr]\biggr\}\nonumber\\*
 &\quad{}-\frac{C_AT_Rn_f}{24}\bigl(55+2\pi^2\bigr)\Biggr\}+O(a_s^3).
 \label{eq:gluon-adjoint-electric-vertex-two-loop}
\end{align}
\begin{align}
 \zeta_{\perp\perp}={}&1+\frac{a_sC_A}{4}\bigl[L(7-3\xi)+1-3\xi\bigr]
 +a_s^2\Biggl\{L^2\left[
 \frac{C_A^2}{96}\bigl(455-180\xi+9\xi^2\bigr)-\frac76C_AT_Rn_f\right]\nonumber\\
 &\quad{}+L\left[\frac{C_A^2}{144}\bigl(1528-621\xi-9\xi^2-96\pi^2\bigr)
 -\frac{91}{18}C_AT_Rn_f\right]\nonumber\\
 &\quad{}+\frac{C_A^2}{3456}\biggl\{
 -2304\pi\bigl[-3+2\Cl(\pi/3)\bigr]
 -9\pi^2\bigl[59+\xi+6\xi^2-256\ln(2)\bigr]\nonumber\\*
 &\qquad{}-11520\ln^2(2)
 +54\xi\Bigl[-186-17\xi+8(\xi-31)\ln(2)\nonumber\\*
 &\qquad\qquad{}-12(\xi-15)\ln(3)+(111+\xi)\Li_2(1/4)\Bigr]\nonumber\\*
 &\qquad{}+8\Bigl[10462+46413\ln(2)-42606\ln(3)-459\Li_2(1/4)\nonumber\\*
 &\qquad\qquad{}+324\Li_3(1/4)+432\zeta(3)\Bigr]\biggr\}
 -\frac{C_AT_Rn_f}{216}\bigl(1207+42\pi^2\bigr)\Biggr\}+O(a_s^3).
 \label{eq:gluon-adjoint-magnetic-vertex-two-loop}
\end{align}
\paragraph{Chromomagnetic field insertion.}
For the operator
$J_F^{\mu\nu}=\bar h_v^{\mathrm F}F_{\perp\perp}^{\mu\nu}h_v^{\mathrm F}$
in eq.~\eqref{eq:fundamental-magnetic-current}, inserted on a fundamental Wilson
line or loop, the vertex factor in eq.~\eqref{eq:gluon-factorization} is
\begin{align}
 \zeta_F={}&1-a_s\Biggl\{L\left[\frac{C_A}{4}(\xi-1)+C_F(\xi-3)\right]
 +\frac{C_A}{4}(1+\xi)+C_F(\xi-1)\Biggr\}\nonumber\\*
 &\hphantom{1}{}+a_s^2\Biggl\{L^2\biggl[
 \frac{C_F^2}{2}(\xi-3)^2+\frac{C_AC_F}{4}(25-7\xi)
 +\frac{C_A^2}{96}(47-24\xi-3\xi^2)\nonumber\\*
 &\qquad\qquad{}-2C_FT_Rn_f-\frac16C_AT_Rn_f\biggr]\nonumber\\
 &\quad{}+L\biggl[C_F^2(\xi-1)(\xi-3)
 +\frac{C_AC_F}{12}(217-54\xi-3\xi^2)
 +\frac{C_A^2}{144}(118-153\xi-27\xi^2)\nonumber\\*
 &\qquad\qquad{}-\frac{20}{3}C_FT_Rn_f-\frac{31}{18}C_AT_Rn_f\biggr]\nonumber\\
 &\quad{}+\frac{C_F^2}{2}(\xi-1)^2
 -\frac{C_AC_F}{48}\biggl\{
 -2140-\pi(96+\pi)+204\xi+(24+\pi^2)\xi^2\nonumber\\*
 &\qquad{}+48(5\xi-106)\ln(2)-216(\xi-24)\ln(3)\nonumber\\*
 &\qquad{}-144(\xi-3)\Li_2(1/4)-72\Li_3(1/4)-96\zeta(3)\biggr\}\nonumber\\
 &\quad{}+\frac{C_A^2}{3456}\biggl\{
 6224+9\pi(384+193\pi)-9(204+\pi^2)\xi-18(27+\pi^2)\xi^2\nonumber\\*
 &\qquad{}+216\bigl[871+2\xi(\xi-11)\bigr]\ln(2)-11520\ln^2(2)\nonumber\\*
 &\qquad{}-648\bigl[238+\xi(\xi-3)\bigr]\ln(3)
 +54\bigl[220+\xi(15+\xi)\bigr]\Li_2(1/4)\biggr\}\nonumber\\
 &\quad{}-\frac{C_FT_Rn_f}{3}\bigl(20+\pi^2\bigr)
 -\frac{C_AT_Rn_f}{216}\bigl(487+6\pi^2\bigr)\Biggr\}+O(a_s^3).
 \label{eq:gluon-fundamental-magnetic-vertex-two-loop}
\end{align}

\Needspace{6\baselineskip}
\paragraph{Chromoelectric field insertion.}
For the operator
$J_E^{\mu\nu}=\bar h_v^{\mathrm F}F_{\parallel\perp}^{\mu\nu}h_v^{\mathrm F}$
inserted on a fundamental Wilson line or loop, the vertex matching factor is
\begin{align}
 \zeta_E={}&1-a_s\Biggl\{L\left[\frac{C_A}{4}(3+\xi)+C_F(\xi-3)\right]
 +\frac{C_A}{4}(1+\xi)+C_F(\xi-1)\Biggr\}\nonumber\\*
 &\hphantom{1}{}+a_s^2\Biggl\{L^2\biggl[
 \frac{C_F^2}{2}(\xi-3)^2+\frac{C_AC_F}{4}(13-3\xi)
 -\frac{C_A^2}{32}(35+\xi^2)\nonumber\\*
 &\qquad\qquad{}-2C_FT_Rn_f+\frac12C_AT_Rn_f\biggr]\nonumber\\
 &\quad{}+L\biggl[C_F^2(\xi-1)(\xi-3)
 +\frac{C_AC_F}{12}(205-42\xi-3\xi^2)
 -\frac{C_A^2}{48}(130+39\xi+9\xi^2)\nonumber\\*
 &\qquad\qquad{}-\frac{20}{3}C_FT_Rn_f+\frac76C_AT_Rn_f\biggr]\nonumber\\
 &\quad{}+\frac{C_F^2}{2}(\xi-1)^2
 -\frac{C_AC_F}{48}\biggl\{
 -2140-\pi(96+\pi)+204\xi+(24+\pi^2)\xi^2\nonumber\\*
 &\qquad{}+48(5\xi-106)\ln(2)-216(\xi-24)\ln(3)\nonumber\\*
 &\qquad{}-144(\xi-3)\Li_2(1/4)-72\Li_3(1/4)-96\zeta(3)\biggr\}\nonumber\\
 &\quad{}+\frac{C_A^2}{384}\biggl\{
 -2096+\pi^2-204\xi
 -\xi\bigl[54\xi+\pi^2(1+2\xi)-48(\xi-11)\ln(2)\bigr]\nonumber\\*
 &\qquad{}-72\bigl[-90+(\xi-3)\xi\bigr]\ln(3)
 -6744\ln(2)+6\bigl[-36+\xi(15+\xi)\bigr]\Li_2(1/4)\biggr\}\nonumber\\
 &\quad{}-\frac{C_FT_Rn_f}{3}\bigl(20+\pi^2\bigr)
 +\frac{C_AT_Rn_f}{24}\bigl(25+2\pi^2\bigr)\Biggr\}+O(a_s^3).
 \label{eq:gluon-fundamental-electric-vertex-two-loop}
\end{align}
Combining the vertex matching factor $\zeta_E$ with the field matching factors gives
\begin{equation}
 \zeta_A(\zeta_{h_v}^{\mathrm F})^2\zeta_E=1+O(a_s^3).
 \label{eq:fundamental-electric-unit-matching}
\end{equation}

\section{Master integrals}
\label{app:masters}
\label{app:master-definitions}

We have rescaled the loop momenta and internal flow times to dimensionless
variables and absorbed the dimensional factors into the coefficients
multiplying the master integrals.

We use $D=4-2\ep$ and
\begin{equation}
 \int_k=\int\frac{\dd^D k}{(2\pi)^D},\qquad
 \int_{k,l}=\int_k\int_l.
 \label{eq:app-master-measure}
\end{equation}
All results below are expanded through the orders in $\ep$ needed for our
calculation.

We write the eikonal propagator as $-\ii/E(q)$, with
\begin{equation}
 E(q)=v\cdot q-\ii0^+,\qquad v^2=1.
 \label{eq:app-master-conventions}
\end{equation}
\paragraph{One-loop integrals.}
The one-loop master integrals are:
\begin{align}
 J_1&=(4\pi)^{D/2}\int_k\frac{e^{-2k^2}}{(k^2)^2}
 =\frac{2^{4-D/2}}{(D-2)(D-4)}\nonumber\\*
 &=-\frac1\ep-1-\ln(2)
 -\left(1+\ln(2)+\frac{\ln^2(2)}{2}\right)\ep+O(\ep^2),
 \label{eq:app-one-loop-radial}\\
 J_2&=-(4\pi)^{D/2}\int_k\frac{e^{-2k^2}}{k^2E(k)^2}
 =\frac{2^{4-D/2}}{D-4}\nonumber\\*
 &=-\frac2\ep-2\ln(2)-\ep\ln^2(2)+O(\ep^2).
 \label{eq:app-one-loop-static}
\end{align}

\paragraph{Two-loop integrals without static propagators.}
\begin{align}
 I_1&=(4\pi)^D\int_{k,l}e^{-2l^2-2(k+l)^2}\nonumber\\*
 &=\frac1{16}+\frac{\ln(2)}{8}\ep
 +\frac{\ln^2(2)}{8}\ep^2+O(\ep^3),
 \label{eq:app-master-I1}\\
 I_2&=(4\pi)^D\int_{k,l}e^{k^2-3l^2-3(k+l)^2}\nonumber\\*
 &=\frac19+\frac{\ln(3)}{9}\ep
 +\frac{\ln^2(3)}{18}\ep^2+O(\ep^3),
 \label{eq:app-master-I2}\\
 I_3&=(4\pi)^D\int_{k,l}\frac{e^{-2(k+l)^2}}{k^2l^2}\nonumber\\*
 &=\frac1{4\ep}+\frac{1+2\ln(2)}{4}
 +\frac{6+\pi^2+12\ln(2)+12\ln^2(2)}{24}\ep+O(\ep^2),
 \label{eq:app-master-I3}\\
 I_4&=(4\pi)^D\int_{k,l}\frac{e^{k^2-3l^2-3(k+l)^2}}{l^2(k+l)^2}\nonumber\\*
 &=\ln(4/3)
 +\bigl[\ln(4/3)\bigl(1+\tfrac12\ln(12)\bigr)+\operatorname{Li}_2(1/4)\bigr]\ep+O(\ep^2),
 \label{eq:app-master-I4}\\
 I_5&=(4\pi)^D\int_0^1\dd u\int_{k,l}
 e^{-(2-u)k^2-u l^2-u(k+l)^2}\nonumber\\*
 &=\frac1{32\ep}+\frac{-7+12\ln(2)-3\ln(3)}{96}\nonumber\\*
 &\hphantom{{}=\frac1{32\ep}}{}+\frac{-14-24\ln(2)+24\ln^2(2)-2\ln(3)-3\ln^2(3)
 -12\operatorname{Li}_2(1/4)}{192}\ep+O(\ep^2),
 \label{eq:app-master-I5}\\
 I_6&=(4\pi)^D\int_0^1\dd u\int_{k,l}
 e^{-(1-u)k^2-(1+u)l^2-(1+u)(k+l)^2}\nonumber\\*
 &=\frac1{24}+\frac{\ln(3)}{32}
 {}+\frac{8-\pi^2+24\ln^2(2)+2\ln(3)+3\ln^2(3)
 +12\operatorname{Li}_2(1/4)}{192}\ep+O(\ep^2),
 \label{eq:app-master-I6}\\
 I_7&=(4\pi)^D\int_{k,l}\frac{e^{-2l^2-2(k+l)^2}}{k^2(k+l)^2}\nonumber\\*
 &=\frac{\ln(2)}{4}
 +\frac{\pi^2+12\ln(2)+24\ln^2(2)}{48}\ep+O(\ep^2).
 \label{eq:app-master-I7}
\end{align}

\vspace{3pt}
\paragraph{Two-loop integrals with static propagators.}
\begin{align}
 I_8&=(4\pi)^D\int_{k,l}
 \frac{e^{-2(k^2+l^2)}}{k^2l^2E(k)E(l)E(k+l)^2}\nonumber\\*
 &=-\frac4\ep+8-4\pi-16\ln(2)+O(\ep),
 \label{eq:app-master-I8}\\
 I_9&=-(4\pi)^D\int_{k,l}\frac{e^{-2(k^2+l^2)}}{(k^2)^2l^2}
 \left[\frac1{E(k)E(-l)}-\frac1{E(k)E(k+l)}\right]\nonumber\\*
 &=\frac1{\ep^2}+\frac{2\ln(2)-2}{\ep}
 +2\ln^2(2)-8\ln(2)+4-4\pi+O(\ep),
 \label{eq:app-master-I9}\\
 I_{10}&=-(4\pi)^D\int_{k,l}
 \frac{e^{-k^2-l^2-(k+l)^2}}{k^2l^2E(k)E(-l)}\nonumber\\*
 &=\frac{5\pi^2}{9}+2\operatorname{Li}_2(1/4)+O(\ep),
 \label{eq:app-master-I10}\\
 I_{11}&=-(4\pi)^D\int_{k,l}
 \frac{e^{-k^2-l^2-(k+l)^2}}{k^2l^2(k+l)^2E(-k)E(l)}\nonumber\\*
 &=-\frac{2\pi^2}{3\ep}-\frac{4\pi^2}{3}
 -\frac{20\pi}{3}\operatorname{Cl}_2(\pi/3)+O(\ep),
 \label{eq:app-master-I11}\\
 I_{12}&=-(4\pi)^D\int_0^1\dd u\int_{k,l}
 \frac{e^{-2k^2-2l^2-2(1-u)k\cdot l}}{k^2l^2E(k)E(-l)}\nonumber\\*
 &=\frac{4\pi}{3}\operatorname{Cl}_2(\pi/3)
 +\frac43\zeta(3)+\operatorname{Li}_3(1/4)+O(\ep),
 \label{eq:app-master-I12}\\
 I_{13}&=-(4\pi)^D\int_0^1\dd u\int_{k,l}
 \frac{e^{-2k^2-2l^2-2(1-u)k\cdot l}}{k^2l^2E(k)E(l)}\nonumber\\*
 &=\frac{8\pi}{3}\operatorname{Cl}_2(\pi/3)
 -\frac43\zeta(3)-\operatorname{Li}_3(1/4)+O(\ep),
 \label{eq:app-master-I13}\\
 I_{14}&=-(4\pi)^D\int_{k,l}
 \frac{e^{-k^2-l^2-(k+l)^2}}{(k^2)^2l^2E(-k)E(l)}\nonumber\\*
 &=\frac1\ep-2+\ln(27/16)
 -\frac{2\pi}{9}(6\sqrt3+5\pi)-4\operatorname{Li}_2(1/4)+O(\ep).
 \label{eq:app-master-I14}
\end{align}
\bibliographystyle{JHEP}
\bibliography{references}
\end{document}